
\raggedbottom
\font\poem=cmbxti10
\def\approx{\simeq}
%
%
\catcode`@=11
\newcount\chapternumber      \chapternumber=0
\newcount\sectionnumber      \sectionnumber=0
\newcount\equanumber         \equanumber=0
\let\chapterlabel=0
\newtoks\chapterstyle        \chapterstyle={\Number}
\newskip\chapterskip         \chapterskip=\bigskipamount
\newskip\sectionskip         \sectionskip=\medskipamount
\newskip\headskip            \headskip=8pt plus 3pt minus 3pt
\newdimen\chapterminspace    \chapterminspace=15pc
\newdimen\sectionminspace    \sectionminspace=10pc
\newdimen\referenceminspace  \referenceminspace=25pc
\def\chapterreset{\global\advance\chapternumber by 1
   \ifnum\the\equanumber<0 \else\global\equanumber=0\fi
   \sectionnumber=0 \makel@bel}
\def\makel@bel{\xdef\chapterlabel{%
\the\chapterstyle{\the\chapternumber}.}}
\def\sectionlabel{\number\sectionnumber \quad }
\def\unnumberedchapters{\let\makel@bel=\relax \let\chapterlabel=\relax
\let\sectionlabel=\relax \equanumber=-1 }
\def\eqname#1{\relax \ifnum\the\equanumber<0
     \xdef#1{{\rm(\number-\equanumber)}}\global\advance\equanumber by -1
    \else \global\advance\equanumber by 1
      \xdef#1{{\rm(\chapterlabel \number\equanumber)}} \fi}

\def\eqn#1{\eqno\eqname{#1}#1}

\def\eqinsert#1{\noalign{\dimen@=\prevdepth \nointerlineskip
   \setbox0=\hbox to\displaywidth{\hfil #1}
   \vbox to 0pt{\vss\hbox{$\!\box0\!$}\kern-0.5\baselineskip}
   \prevdepth=\dimen@}}
%

%

%

%
%
\newcount\fcount \fcount=0
\def\ref#1{\global\advance\fcount by 1
  \global\xdef#1{\relax\the\fcount}}
\def\refs{\noindent \hangindent=5ex\hangafter=1}
\def\np{\vfill\eject}
%
%
\def\today{\number\year\space \ifcase\month\or 	January\or February\or
	March\or April\or May\or June\or July\or August\or September\or
	October\or November\or December\fi\space \number\day}
%
%
\def\spose#1{\hbox to 0pt{#1\hss}}
\def\simlt{\mathrel{\spose{\lower 3pt\hbox{$\mathchar"218$}}
     \raise 2.0pt\hbox{$\mathchar"13C$}}}
\def\simgt{\mathrel{\spose{\lower 3pt\hbox{$\mathchar"218$}}
     \raise 2.0pt\hbox{$\mathchar"13E$}}}
\def\simpropto{\mathrel{\spose{\lower 3pt\hbox{$\mathchar"218$}}
     \raise 2.0pt\hbox{$\propto$}}}
\def\frac#1/#2{\leavevmode\kern.1em
 \raise.5ex\hbox{\the\scriptfont0 #1}\kern-.1em
 /\kern-.15em\lower.25ex\hbox{\the\scriptfont0 #2}}
%
%
\newif\ifmathmode
\def\mathflag#1${\mathmodetrue#1\mathmodefalse$}
\everymath{\mathflag}
\def\displayflag#1$${\mathmodetrue#1\mathmodefalse$$}
\everydisplay{\displayflag}
\mathmodefalse

\font\tenmmib=cmmib10
\font\sevenmmib=cmmib10
\font\fivemmib=cmmib10
\newfam\bmitfam\textfont\bmitfam=\tenmmib
\scriptfont\bmitfam=\sevenmmib\scriptscriptfont\bmitfam=\fivemmib
\def\fixfam#1{\ifmathmode
                 {\ifnum\fam=\bffam
                    {\fam\bmitfam#1}\else
                    {\fam1#1}\fi}\else
                 {\ifnum\fam=\bffam
                    {$\fam\bmitfam#1$}\else
                    {$\fam1#1$}\fi}\fi}

\def\alpha{\fixfam{\mathchar"710B}}
\def\beta{\fixfam{\mathchar"710C}}
\def\gamma{\fixfam{\mathchar"710D}}
\def\delta{\fixfam{\mathchar"710E}}
\def\epsilon{\fixfam{\mathchar"710F}}
\def\zeta{\fixfam{\mathchar"7110}}
\def\eta{\fixfam{\mathchar"7111}}
\def\theta{\fixfam{\mathchar"7112}}
\def\iota{\fixfam{\mathchar"7113}}
\def\kappa{\fixfam{\mathchar"7114}}
\def\lambda{\fixfam{\mathchar"7115}}
\def\mu{\fixfam{\mathchar"7116}}
\def\nu{\fixfam{\mathchar"7117}}
\def\xi{\fixfam{\mathchar"7118}}
\def\pi{\fixfam{\mathchar"7119}}
\def\rho{\fixfam{\mathchar"711A}}
\def\sigma{\fixfam{\mathchar"711B}}
\def\tau{\fixfam{\mathchar"711C}}
\def\upsilon{\fixfam{\mathchar"711D}}
\def\phi{\fixfam{\mathchar"711E}}
\def\chi{\fixfam{\mathchar"711F}}
\def\psi{\fixfam{\mathchar"7120}}
\def\omega{\fixfam{\mathchar"7121}}
\def\varepsilon{\fixfam{\mathchar"7122}}
\def\vartheta{\fixfam{\mathchar"7123}}
\def\varpi{\fixfam{\mathchar"7124}}
\def\varrho{\fixfam{\mathchar"7125}}
\def\varsigma{\fixfam{\mathchar"7126}}
\def\varphi{\fixfam{\mathchar"7127}}

\def\Bv{{\bf v}}
\def\Bk{{\bf k}}

\def\Bg{{\bf \gamma}}
\def\part#1;#2 {\partial#1 \over \partial#2}
\def\deriv#1;#2 {d#1 \over d#2}
\def\G {\gamma}
\def\D {\Delta}
\def\E {\eta}
\def\Oo {\Omega_0}
\def\Mc {{\bf j}}
\def\neav {\bar n_e}
\def\X{{\bf x}}
\def\Ec {\eta_*}
\def\Da {\Delta^{(1)}}
\def\Db {\Delta^{(2)}}
\def\da {\delta^{(1)}}
\def\db {\delta^{(2)}}
\def\Le {{\cal R}}
\def\drat {{\cal D}}
\def\eden {{\cal E}}
\def\gnnp {g(\E,\E')}
\def\gnn {\dot \tau (\E)}
\def\va {v^{(1)}}

\def\vd {v \delta}
\def\P{{\bf p}}
\def\Q{{\bf q}}
\def\V{{\bf v}}
\def\K{{\bf k}}
\def\cos{{\rm cos}}
\def\sin{{\rm sin}}
\def\exp{{\rm exp}}
\def\ct {\cos \theta}
\def\cst {\cos^2 \theta}
\def\e {{\rm e}}
\def\Cy {Compton-$y$}
\def\opc{\left( 1 + \cos^2 \beta \right)}
\def\cosb{\cos \beta}
\def\del{\delta(p-p')}
\def\ddel{\left[ {\partial \over {\partial p'}} \del \right]}
\def\dddel{\left[ {\partial^2 \over {\partial p'^2}} \del \right]}
\def\fmf{F_1(\X,\P,\P')}
\def\ffff{F_2(\X,\P,\P')}
\def\fof{F_3(\X,\P,\P')}
\def\pmp{(\P-\P')}
\def\bomb{\cosb (1-\cosb)}
\def\pv{\V \cdot \P}
\def\ppv{\V \cdot \P'}
\def\pq{\Q \cdot \P}
\def\ppq{\Q \cdot \P'}
\def\df{{\partial f \over \partial p}}
\def\dfo{{\partial f_0 \over \partial p}}
\def\ddf{{\partial^2 f \over \partial p^2}}
\def\eg{{\it e.g.,\ }}
\def\etal{{\it et al.}}
\def\ie{{\it i.e.,\ }}

\def\eqnoh#1{\eqno\hbox{#1}}
\def\MNRAS{Mon. Not. Roy. Astron. Soc.}
\bigskip
\unnumberedchapters


\font\titlefont=cmbx10 at 14.4truept
\font\namefont=cmr12
\font\addrfont=cmti12
\font\rmtwelve=cmr12

\newbox\abstr
\def\abstract#1{\setbox\abstr=
    \vbox{\hsize 5.0truein{\par\noindent#1}}
    \centerline{ABSTRACT} \vskip12pt
    \hbox to \hsize{\hfill\box\abstr\hfill}}

\def\today{\ifcase\month\or
        January\or February\or March\or April\or May\or June\or
        July\or August\or September\or October\or November\or December\fi
        \space\number\day, \number\year}

\def\author#1{{\namefont\centerline{#1}}}
\def\addr#1{{\addrfont\centerline{#1}}}

{ 
\nopagenumbers
\rmtwelve

\vsize=9 truein
\hsize=6.5 truein
\raggedbottom
\baselineskip=16pt

\line{\hfil CfPA-TH-93-15}
\line{\hfil $28^{\rm th}$ May 1993}
\vskip2truecm
\centerline
{\titlefont REIONIZATION}
\centerline
{\titlefont AND COSMIC MICROWAVE BACKGROUND DISTORTIONS:}
\centerline
{\titlefont A COMPLETE TREATMENT OF}
\centerline
{\titlefont SECOND ORDER COMPTON SCATTERING
\footnote{$^\dagger$}{\rm Submitted to Physical Review D}}

\nobreak
  \vskip 1.0truecm
  \bigskip
  \author{Wayne Hu, Douglas Scott \& Joseph Silk}
  \smallskip
  \addr{Departments of Physics and Astronomy}
  \addr{and Center for Particle Astrophysics,}
  \addr{University of California, Berkeley, California  94720}
  \bigskip

\noindent{\rm
The ionization history of the universe provides a major source of ambiguity in
constraining cosmological models using small angular scale microwave
background anisotropies.  To clarify these issues,
we consider a complete treatment of Compton scattering to second
order, an approach which may be applicable
to other astrophysical situations.  We find that only the ${\cal O} (v)$
Doppler effect, and the ${\cal O} (v\delta)$ Vishniac effect are important
for recent last scattering epochs; the ${\cal O} (v^2)$ Doppler effect
is not significant on any angular scale, and other higher-order effects
are completely negligible.  However the ${\cal O}(v^2)$ effect does lead to
\Cy\ distortions, which although generally below current constraints,
set an unavoidable minimum level in reionization models.
We consider the small-angle approximation for the Vishniac effect in
detail, and show several improvements over previous treatments, particularly
for low $\Omega_0$.   For standard cold dark matter models, the effect
of reionization is to redistribute the anisotropies to arcminute scales;
late reionization leads to partially erased primary fluctuations and a
secondary contribution of comparable magnitude.
Using recent anisotropy limits from the ATCA experiment, we set new constraints
on baryonic dark matter models.   Stronger constraints are imposed (in second
order) upon models with higher Hubble constant, steeper $n$, and higher
density.  These limits depend on the specific
ionization history assumed, but the factor gained by lowering the
ionization fraction is generally small, and may be tested by currently-planned
experiments on arcminute scales.}
\bigskip
\vfill
\centerline{\poem Mingled and merged, densely sprouting}

\centerline{\poem In the primeval mass, there is no shape}

\centerline{\poem Spreading and scattering, leaving no trail behind}

\centerline{\poem In the darkness of its depths there is no sound}

\centerline{\poem --Chuang-tzu}
\bigskip
\eject
}  

\goodbreak

\centerline {\bf I. INTRODUCTION}
\smallskip
Temperature fluctuations in the
cosmic microwave background (CMB) are a direct probe of
density perturbations at $z\simeq1100$.  Therefore measurements or upper limits
on such fluctuations can be used to constrain cosmological models for the
evolution of structure.  However, there is a possible loop-hole at small
angular scales, since reionization could make the last scattering epoch
more recent, and erase the small-scale fluctuations,
\eg \ref\HKR [\HKR].
This
possibility requires significant levels of ionization back to $z\gg10$, to
reach optical depth unity at $z\ll1100$, which is certainly feasible in
some models.  With the recent detection of temperature anisotropies at large
angular scales by Differential Microwave Radiometer (DMR) on
the Cosmic Background Explorer (COBE) satellite
\ref\COBE [\COBE], 
the constraining of cosmological models has entered a new phase of
precision, with application to intermediate angular scale anisotropy
experiments, \eg \ref\Gaier [\Gaier].
It is becoming increasingly important to know how much leeway
reionization scenarios can give for small angular scale limits.  In
this spirit, we have attempted to systematically consider the effects
of Compton scattering on CMB photons.

Primary fluctuations arising from the recombination epoch are erased on
scales smaller than that subtended by the horizon at the new last
scattering epoch.  However, secondary fluctuations will be generated on this
new surface of last scattering, mainly due to Doppler shifts among the
scatterers.  These secondary fluctuations are ${\cal O}(v)$, except that
there is a partial cancellation of blueshifts and redshifts through a given
overdense (or underdense) region.  The secondary fluctuations were therefore
thought to be small until Ostriker \& Vishniac
\ref\OV [\OV]
pointed out that a
second order contribution [of ${\cal O}(v\delta)$, the so-called Vishniac
term],
which does not
suffer from the cancellation, could be larger than the first order term.
In practice the second order term
is of the same order of magnitude as the degree-scale
primary fluctuations which have been erased, but on arcminute scales
and below.

Use of reionization to avoid small-scale $\Delta T/T$
limits therefore depends on a calculation of these second-order
contributions for the particular power spectrum and ionization history
being considered.
Detailed calculations for some models were performed by Vishniac
\ref\Vishniac [\Vishniac] and Efstathiou \ref\Efstathiou [\Efstathiou].
However, the Vishniac term is only one of a
number of possible second-order terms.  Although it has been assumed that
this term dominates over any others, this has not been demonstrated.  Here
we show that it is reasonable to neglect all other second order terms.
Consideration of secondary anisotropies is therefore calculated to reasonable
accuracy (in the linear regime) if the first order and  Vishniac
contributions to $\Delta T/T$ are evaluated. To arrive at
this result, we set up in \S II a methodology which finds all
second-order Compton scattering terms.  Along the way, we uncover
a number of physical effects which have not previously been described, \eg
minimal spectral distortions required in reionization scenarios
independent of thermal history.
Our general results, detailed in Appendices A and B,
may also be of use for a wider range of problems.
In \S III, we discuss the second order contributions to anisotropy and
significantly improve the approximations of
Efstathiou [\Efstathiou].  Predictions for
the cold dark matter (CDM) and baryonic dark matter with isocurvature
fluctuations (BDM, also known
as primordial baryon isocurvature) scenarios
are computed in \S IV.
Recent limits on arcminute scale fluctuations from the Australian
Telescope Compact Array (ATCA)
\ref\Subrahmanyan [\Subrahmanyan]
place strong constraints
on BDM models.
Other studies of the effects of reionization on the microwave background
[\Vishniac, \Efstathiou]
have tended to concentrate on the simplest case of a universe in which there
has been no recombination.  Here we have also considered the effects of
more realistic ionization histories.
\bigskip
\goodbreak
\centerline{\bf II. THE BOLTZMANN EQUATION }
\centerline{\bf FOR SECOND ORDER COMPTON SCATTERING}
\medskip
\noindent
A. GENERAL FORMALISM
\smallskip

The Boltzmann equation in general is given by
$$
{\part f;t } + {\part f;x^i }{\deriv x^i;t }
+ {\part f;\gamma_i } {\deriv \gamma_i;t }
+ {\part f;p_0 }{\deriv p_0;t } = C(x,p),
\eqn\eqnBoltFull
$$
where  $f$ is the photon occupation number,
$\gamma_i$ are the direction cosines for
a photon of 4-momentum $p$, and
the expression $C(x,p)$ is the usual collision
term.  Latin indices range from 1--3, and we have employed the implicit
summation convention.
There are two distinct classes of second order Boltzmann equations
that we might consider.
The left hand side of equation \eqnBoltFull\ may be expanded to
second order in metric perturbations; whereas the right hand side,
in the case of Compton scattering,
may be expanded to second order
in the energy transfer from collisions.
In Appendix A, we carry through a full derivation of all
second order gravitational terms due to metric perturbations
 in the synchronous gauge.
However, in this paper we are primarily interested
in the effects of Compton scattering which are constrained by causality
to manifest themselves on small scales where gravitational effects are
unimportant.
Thus we take the zeroth order approximation of equation
\eqnBoltFull\
with respect to the metric fluctuations and find
$$
{\part f; t}
+ {\part f;x^i }{\deriv x^i;t }
- {1 \over a}{da \over dt} p_0 {\part f; p_0} = C(x,p).
\eqn\eqnBolt
$$
The third term on the left merely expresses the
cosmological redshift of the photon
energy, $p_0 \propto a^{-1}$, where $a(t)$ is the usual scale factor.
In the
homogeneous and
isotropic limit, the second term in equation \eqnBolt\ vanishes.
Spectral distortions
in the early universe
are usually computed in this approximation (see part B).
In \S III, we calculate the effects of dropping the assumption of
homogeneity.
If, on the
other hand, we are only concerned with temperature and
not spectral distortions, we may integrate over momentum $p$.  The third term
may thus be eliminated since temperature and energy redshift in the same
manner.

The goal is to derive the collision term for Compton scattering,
$\gamma(p) + e(q)
\leftrightarrow \gamma(p') + e(q')$,
to second order in the small energy transfer due to scattering.
Note that we are performing calculations in the {\it linear} regime.
Thus, the
results are of interest only when the effects of the first order terms
suffer
cancellation, as in the case of a thick last scattering surface.
The lowest order term is then of second order.  Third and
higher order effects are therefore negligible as long as the second
order term is not cancelled.

Our approach may be of more
general interest since it provides a coherent framework
for all Compton scattering effects,
be they spectral distortions or anisotropies.
In the proper limits, the equation derived below reduces to
familiar equations and effects, \eg the
Kompaneets equation (Sunyaev-Zel'dovich effect), linear Doppler
equation
(Vishniac effect).  Furthermore, new truly second order effects
such as the ${\cal O}(v^2)$ quadratic Doppler effect are obtained.
These effects predict distortions well below
the observational limits today.  In principle, however, the fact that
they may mix both anisotropies and spectral distortions makes them
distinguishable.

We make the following assumptions in deriving the equations:
(1) the Thomson limit applies, \ie the fractional energy transfer $\delta p/p
\ll 1$ in the rest frame of the
background radiation; (2) the radiation is unpolarized and remains so; (3)
the density of electrons is low so that Pauli suppression terms may be
ignored; and (4) the electron distribution is thermal about some bulk flow
velocity $\V$.
Approximations (1), (3), and (4) are valid for most situations of cosmological
interest.  The approximation regarding polarization could be dropped
to give coupled equations for perturbations in the total and polarized
components
\ref\BE [\BE].
Polarization
perturbations are typically an order of magnitude smaller than temperature
distortions
\ref\KaiserA [\KaiserA].
Thus the contribution from polarization to the evolution of
{\it temperature}
distortions is a small effect, as the microwave background never
generates a significant polarization.  For calculations on the generation
of polarization in second order theory, see [\Efstathiou].
\smallskip
The collision term may in general be expressed as
\ref\Bernstein [\Bernstein]
$$
\eqalign {
C(\X,\P) &= {1 \over {2E(p)}} \int Dq Dq' Dp' (2\pi)^4 \delta^{(4)}
(p+q-p'-q') |M|^2 \cr
&\qquad \times \left\{ g(\X,\Q')f(\X,\P') \left[ 1 + f(\X,\P) \right] -
g(\X,\Q)f(\X,\P)\left[ 1 + f(\X,\P') \right] \right\}, }
\eqn\eqnCGen
$$
where $|M|^2$ is the Lorentz invariant matrix element, $f(\X,\P)$ is the
photon distribution function, $g(\X,\Q)$ is the electron distribution function
and
$$
Dq = { d^3 q \over {(2\pi)^3 2E(q)}}
$$
is the Lorentz invariant phase space element.  The terms in \eqnCGen\ which
contain the distribution functions are just the contributions from
scattering into and out of the momentum state $\P$ including
stimulated emission effects.

We will assume that the electrons are thermally distributed about some
bulk flow velocity $\V$,
$$
g(\X,\Q) = (2\pi)^3 n_e (2\pi mT_e)^{-3/2} \exp \left\{ {- [\Q-m\V(\X)]^2
\over 2mT_e } \right\} ,
\eqn\eqnEDist
$$
where $m$ is the electron mass, and we employ units with $c=\hbar=k=1$.
Expressed in the rest frame of the electron,
the matrix element for Compton scattering summed over polarization is
given by
\ref\MS [\MS]
$$
|M|^2 = 2 (4\pi)^2 \alpha^2 \left[ {\tilde p' \over \tilde p} +
{\tilde p \over \tilde p'}
-  \sin^2 \tilde \beta \right],
\eqn\eqnMCompton
$$
where the tilde denotes quantities in the rest frame of the electron,
$\alpha$ is the fine structure constant,
and
$\cos  \tilde \beta = {\tilde \Bg \cdot \tilde \Bg'}$
is the scattering angle.
Note that here and below we define $p \equiv p_0=|\P|$.
 Of course, the matrix element must be
expressed in terms of the corresponding quantities in the
frame of the radiation for calculation purposes [see equation (B--1)].

The result of integrating over the electron momenta can be written
$$
C(\X,\P) = n_e \sigma_T \int dp' {p' \over p}
\int {d\Omega' \over {4\pi}} {3 \over 4}
\left[ C_0 + C_{p/m} + C_v + C_{vv} + C_{T_e/m} +
C_{vp/m} + C_{(p/m)^2}  \right],
\eqn\eqnCImplicit
$$
where we have kept terms to second order in $\delta p/p$.
The explicit expressions for the quantities in equation \eqnCImplicit\
 are given in equations
(B--4) and (B--5) of Appendix B and are discussed in turn below.
This equation may be
considered as the source equation for all first and second order
Compton scattering effects.  However, in most cases of interest only
a few of these terms will ever contribute.  We have therefore written
the collision equation implicitly and expressed it in terms of the
scaling behavior of the contributing elements.

The expansion of the Compton collision term to second order in
$\delta p /p$  has actually
involved  {\it several} small quantities.
It is worthwhile to
compare these terms.
The quantity
$T_e/m$ characterizes the kinetic energy of the electrons and is to
be compared with $p/m$ or essentially $T/m$ where $T$ is the
temperature of the photons.  Before a redshift  $8.0 (\Omega_0h^2)^{1/5}
x_e^{-2/5}$, where $x_e$ is the ionization fraction (this corresponds to
$z \simgt 500(\Omega_Bh^2)^{2/5}$ for standard recombination),
the tight coupling between photons and electrons via Compton scattering
requires these two temperatures to be comparable.
At lower redshifts,
it is possible that
$T_e \gg T$, which produces distortions in the radiation via the
Sunyaev-Zel'dovich (SZ) effect as discussed further below.   For the
reionization scenarios
which we consider, these two temperatures will typically still be
comparable at last scattering,
 $z_*\approx 50$ with $T/m \approx 5 \times 10^{-10} (1+z_*)$.
The SZ effect, however, will play a role at lower redshifts
even in these scenarios.
Note that the term $T_e/m$ may also be thought of as the
average thermal velocity squared $\langle v^2_{th} \rangle = 3T_e/m$.
This is to be compared with the bulk velocity squared $v^2$ and
will depend on the specific means of ionization.  The bulk
velocity is related to the fractional overdensity $\delta =\delta \rho /\rho$
by the continuity equation.  On scales much smaller than the horizon,
$v \ll \delta$ in linear theory.

It is appropriate at this point to examine the physical significance
and qualitative features of
each of the sources in the collision term of equation \eqnCImplicit:
\medskip
\goodbreak
\noindent {\bf 1. $C_0$: Anisotropy Suppression.}
\smallskip
In the absence of electron motion, there is no preferred direction.
Thus to zeroth order, scattering makes the radiation distribution
more isotropic.
The $C_0$ term given
by equation (B--4) of Appendix B indeed equalizes the distribution
function over all directions via
scattering into and out of a given mode.
It is therefore responsible for the
suppression of primordial anisotropies by reionization.
It can also significantly affect the regeneration of
{\it inhomogeneities}
when ordinary contributions are suppressed, as in the case of
thick last scattering surfaces (see \S IIIC).  Recall that an
inhomogeneity on the last scattering surface becomes an anisotropy
in the CMB today.
\medskip
\goodbreak
\noindent {\bf 2. $C_v$ and $C_{vv}$: Linear and Quadratic Doppler Effect.}
\smallskip
Aside from the small electron recoil (see 3.),
the kinematics of Thomson scattering
require that no energy be transferred in the rest frame of the electron
\ie $\tilde p' = \tilde p$.
Nevertheless, the transformation into the background frame
induces a Doppler shift, so that
$$
{\delta p \over p} = {{1 - \V \cdot \Bg} \over
{1 - \V \cdot \Bg'}} - 1 = {\V}\cdot (\Bg' - \Bg)
+ (\V \cdot \Bg')\V \cdot (\Bg' - \Bg) + {\cal O}(v^3).
\eqn\eqnDoppler
$$
Notice that in addition to the usual linear term in $v$, there is also
a term quadratic in $v$.

To gain physical insight into these Doppler effects,
we can
approximate the photons as isotropic and
neglect the angular dependence of Thomson scattering,
(we will postpone
discussion of the precise effects until \S III). Averaging over the incoming
direction $\Bg$, we obtain
$$
\left< {\delta p \over p} \right>
\approx {\V}\cdot \Bg'
+ (\V \cdot \Bg')^2.
\eqn\eqnDoppAvga
$$
Thus the linear Doppler effect introduces an energy shift whose sign
depends on direction; whereas the quadratic Doppler effect, which
is positive
definite, always gives rise to a blueshift of the photons.

Now let us consider the case where there are many scattering blobs
so that the directions of the electron velocities are in effect randomized.
In this case, the net linear effect vanishes and
$$
\left< {\delta p \over p} \right>
\approx {1 \over 3} \left< v^2 \right>,
\eqn\eqnDoppAvgb
$$
\ie since redshifts and blueshifts cancel, there is no net
transfer of energy to the photons to first order in $v$.
The quadratic Doppler term however
represents a net energy gain of
$$
\D \equiv {\delta \eden \over \eden} =
{4 \over 3}  \langle v^2 \rangle ,
\eqn\eqnDDoppler
$$
since the energy density $\eden \propto T^4$,
and the Doppler shift of
a blackbody is a blackbody with $T = T_0(1 + \delta p / p)$.
Here, we assume for simplicity that all
the photons in the spectrum scattered once.

These effects are not wholly equivalent to a uniform
Doppler shift.
In averaging over angles above,
we have really superimposed many Doppler shifts
for individual scattering events.
Therefore the resulting
spectrum is a superposition of blackbodies with a range of temperatures
$\Delta T /T = {\cal O}(v)$.  Zel'dovich, Illarionov, \& Sunyaev
\ref\ZIS [\ZIS]
have shown that this sort of superposition leads to spectral distortions
of the \Cy\ type with $y = {\cal O}(v^2)$.
Thus, spectral distortions
have a quadratic dependence on $v$ and may be considered as part
of the quadratic Doppler effect.

In summary,  the linear Doppler effect is primarily cancelled after
many scatterings.  Nevertheless, some residual effects remain due to the
evolution of the electron velocities and densities during the last
scattering epoch
\ref\KaiserB [\KaiserB]
and the increased probability of
scattering in an overdense region [\OV].
The quadratic Doppler
effect leads to an average net energy increase of $\D \approx
4\frac/3 \tau \langle v^2
\rangle$,
and a spectral distortion of the \Cy\ type.
Here $\tau =\int n_e \sigma_T dt$ is the
optical depth, which gives the fraction of photons scattered if $\tau \ll 1$.
We will now
see that the quadratic Doppler effect for bulk flows is in fact entirely
equivalent to the
Sunyaev-Zel'dovich effect for thermal motions.

\medskip
\goodbreak
\noindent {\bf 3. $C_{T_e/m}$ and $C_{p/m}$: Thermal Doppler Effect and
Recoil.}
\smallskip
Of course, we have artificially separated out the bulk and thermal components
of the electron velocity.  The thermal velocity leads to a quadratic
Doppler effect exactly as described above if we
make the replacement $\langle v^2 \rangle
\rightarrow \langle v^2_{th} \rangle = 3T_e/m$ [\eg in equation \eqnDDoppler].
For an isotropic
distribution of photons, this
leads to the familiar Sunyaev-Zel'dovich (SZ) effect
\ref\SZ [\SZ].  The SZ effect can therefore be understood as the
second order spectral distortion and energy transfer due to the superposition
of Doppler shifts from individual scattering events off electrons in
thermal motion.
This can
also be naturally interpreted macrophysically:
hot electrons transfer energy to
the photons via Compton scattering.  Spectral distortions result
since low energy photons are shifted upward in frequency, leading to
the Rayleigh-Jeans depletion and the Wien tail enhancement characteristic
of \Cy\ distortions.  The fractional energy change due to
scattering is
$$
\D \approx 4y \approx {4 \over 3} \tau \langle v^2_{th} \rangle.
\eqn\eqnDy
$$
Note that the thermal and quadratic
Doppler effects are wholly equivalent in this sense.

Even though this is a net energy gain, the effective temperature
in the Rayleigh-Jeans region is suppressed due to photon depletion
(up-scattering),
$$
\left({\Delta T \over T} \right)_{RJ,y} = -2 y \approx - {\D \over 2}.
\eqn\eqnTDy
$$
This is to be distinguished from the linear Doppler effect which provides
a frequency independent shift in temperature
given by $\Delta T / T \approx \D/4$ ({\it c.f.} \S\S IIA2).

If the photons have energies comparable to the electrons (\ie the
electron and photon temperatures are nearly equal), there is also a
significant
correction due to the recoil of the electron.  The
scattering kinematics tell us that
$$
{\tilde p' \over \tilde p} =
\left[ 1 + {\tilde p \over m}(1 - \cos \tilde \beta)
\right]^{-1}.
\eqn\eqnRecoil
$$
Thus to lowest order, the recoil effects are ${\cal O}(p/m)$.  Together with
the thermal Doppler effect, these terms form the familiar Kompaneets
equation
\ref\Kompaneets [\Kompaneets]
in the limit where the radiation is isotropic
(as described in part B).
The combination of thermal Doppler shift and recoil drive the photons
to  a Bose-Einstein distribution of temperature $T_e$.
A blackbody distribution cannot generally
be established since Compton scattering
requires conservation of the photon number.

\medskip
\goodbreak
\noindent {\bf 4. $C_{vp/m}$ and $C_{(p/m)^2}$: Higher Order Recoil Effects.}
\smallskip
These terms represent the next order in corrections due to the recoil
effect.
In almost all cases, they are
entirely negligible.  Specifically, for the CMB, when electron bulk flow
effects become significant
$(p/m) \simlt {\cal O}(v^2)$.  Furthermore, since there
is no cancellation in the $C_{p/m}$ term, $C_{(p/m)^2}$ will
never produce the dominant effect.
We will hereafter drop these terms in our consideration.
\bigskip
\goodbreak
\noindent
B. THE ISOTROPIC LIMIT AND SPECTRAL DISTORTIONS
\smallskip
The most interesting case to consider is when the radiation is nearly
isotropic.  For an initially isotropic radiation field,
only the Doppler effects, which have a preferred direction along the
velocity of the electron, can generate an anisotropy.  However, as
we have seen, the first order contributions due to the linear Doppler effect
tend to cancel.
Furthermore, to lowest order, scattering reduces anisotropies.
It is therefore reasonable to consider the case
where the anisotropy itself is at most second order.
We will justify this more fully in \S III.
The general results of Appendix B
can be employed to study the effects of Compton scattering on highly
anisotropic radiation.  For example, equation (B--4) contains the anisotropic
generalization of the Kompaneets equation which may be useful in other
astrophysical settings.

Under the additional assumption of near isotropy,
the collision term becomes more
tractable and transparent:
$$
\eqalign{
C(\X,\P) &= n_e \sigma_T \Bigg\{ \left[ f_0-f
+ {1 \over 2}P_2(\mu)f_2 \right] - \Bg \cdot \V
p \dfo + \Bigg( \left[ (\Bg \cdot \V)^2 + v^2
\right]
p\df \cr
&\qquad + \left[ {3 \over 20} v^2  +  {11 \over 20} (\Bg \cdot \V)^2
 \right] p^2 \ddf
\Bigg)
+ {1 \over mp^2} {\partial \over \partial p} \left[ p^4 \left\{
T_e \df + f(1+f) \right\} \right]  \Bigg\}, }
\eqn\eqnCIso
$$
where
$$
f_l= \int {d\Omega' \over 4\pi} P_l(\mu') f
\eqn\eqnLegendre
$$
are the Legendre
moments of the distribution function with respect to the
axis $x_3$, $\mu = \Bg \cdot \hat \X_3$.
We assume that for plane wave situations there is an
azimuthal symmetry about the direction of the wavevector $\hat \X_3$ since
to lowest order the source term possesses this symmetry (see \S IV).

The first term in equation \eqnCIso\
just represents the zeroth order suppression
and can be found in the standard treatments in the literature, \eg
\ref\Bond [\Bond].
In the limit of complete isotropy of the radiation field, it vanishes
identically.  The second and third terms represent the linear
and quadratic Doppler effects respectively.  The final term is the
usual Kompaneets equation.

Notice that in the limit of many scattering
blobs, we can average over the direction of  the electron velocity.
The first order linear Doppler effect
primarily cancels in this case. Assuming complete isotropy and
ignoring the residual contribution
of the linear Doppler effect,
we can reduce equation \eqnCIso\ to
$$
C(\X,p) = n_e \sigma_T
\Bigg\{{\langle v^2 \rangle \over 3} {1 \over p^2}
{\partial \over \partial p} \left[p^4 \df \right] +
 {1 \over mp^2} {\partial \over \partial p} \left[ p^4 \left\{
T_e \df + f(1+f) \right\} \right]  \Bigg\}.
\eqn\eqnCIsoAvg
$$
Under the replacement $\langle v^2_{th}\rangle = 3T_e/m \rightarrow v^2$,
the SZ (thermal Doppler) portion of the Kompaneets equation
and quadratic Doppler equation
have the same form.
We also regain the factor of 1/3 described in
\eqnDoppAvgb.
Thus, spectral distortions due to bulk flow have exactly the same form
as SZ distortions and can be characterized by the \Cy\ parameter
given in its full form by
$$
y = \int n_e \sigma_T \left[ {1 \over 3} \langle v^2(t) \rangle
+ {T_e - T \over m} \right] dt.
\eqn\eqnY
$$
A full analysis of the
Boltzmann equation analogous to that performed in \S III yields this
result as well.
The appearance of the photon temperature $T$ in equation \eqnY\
is due to the recoil terms in the Kompaneets
equation [{\it c.f.} equation \eqnDy].

As was noted by
Zel'dovich \& Sunyaev
\ref\ZS [\ZS]
the observational constraints
on the \Cy\ parameter in conjunction with other spectral limits can rule out
some reionization histories independently of the detailed thermal history
of the models.  However, Bartlett and Stebbins
\ref\BS [\BS]
show
that for a low $\Omega_B $ universe as implied by nucleosynthesis, no
such model independent constraints exist.  Furthermore, for the
case where the electron temperature and photon temperature are
equal, there is no SZ effect as one can clearly see from equation \eqnY.
Therefore, the bulk flow effect places a lower limit on spectral distortions
due
to any given reionization history.

For example in  an $\Oo=1$ universe with full ionization
up to a redshift $z_i$
and $T_e = T$,
$$
y \approx 4.1 \times 10^{-2} (\Omega_B h)
\left[ (1+z_i)^{1/2}-1 \right] \langle v_0^2 \rangle,
\eqn\eqnYSecond
$$
where $\langle v_0^2 \rangle$
is the (spatially) averaged square of the velocity today.
Here and throughout the paper $\Omega_B$ refers to the fraction
of the critical density {\it of the intergalactic medium} in baryons;
it does not include baryonic matter in compact objects.
We have assumed
linear growth of velocities in an $\Omega_0 =1$ universe,
$$
v(k,\E) =  i {\dot \delta \over k}
={2i \delta \over k\E} \propto \E \qquad [\Oo=1],
\eqn\eqnVlinear
$$
where the dot denotes a derivative with respect to
conformal time $\E = \int dt/a$.
Given a power spectrum $P(k)$ for the baryon density fluctuations,
we can compute the average velocity as
$$
\langle v_0^2 \rangle = {2V_x \over (\pi \eta_0)^2 } \int dk P(k),
\eqn\eqnVavg
$$
assuming ergodicity.  Here $V_x$ is the volume
in which the universe is assumed to be periodic.
The result for the standard cold dark matter (CDM) scenario,
normalized to the COBE measurement [\COBE], is therefore
$v_{0, rms} = 1100$ km/s assuming that the
baryons follow the dark matter distribution (see \S V for details).
Note that this value is rather large, since
COBE requires a bias factor $b \approx 1$ which produces excessive
small scale velocities.  For comparison purposes, the sun's velocity
with respect to the CMB is $365\pm18$ km/s whereas the velocity of
the local group is $622\pm20$ km/s
\ref\Smoot [\Smoot].
The upper limit on the \Cy\ distortion from the Far Infrared Absolute
Spectrophotometer (FIRAS)
\ref\Mather [\Mather] is
$y < 2.5 \times 10^{-5}$.  Therefore it is appropriate to reexpress equation
\eqnYSecond\ as
$$
z_{max} \approx 3.0 \times 10^{3} \left( {y \over 2.5 \times 10^{-5}} \right)^2
\left( {1000\ {\rm km/s} \over v_{0,rms}} \right)^4 {1 \over (\Omega_B h)^2},
\eqn\eqnYconstraint
$$
where  $z_{max}$ is the maximum redshift that the electrons can be fully
ionized.

However our assumption that the electron density
follows the dark matter is invalid  at high redshifts.
At redshifts $z > z_{drag}
\approx 160 (\Oo h^2)^{1/5}x_e^{-2/5}$, Compton drag on the electrons
and protons prevent matter from falling into the dark matter wells.
Thus, \eqnYconstraint\ will not give
a constraint if $z_{max} > z_{drag}$.
Correspondingly, the prediction for CDM is $y(z_{drag})
\approx 3 \times 10^{-7}$, almost two orders of magnitude below the current
limits.
Almost certainly foreground contamination from hot
clusters will make this effect unobservable.  Estimates employing
cluster modeling yield
$y \approx 2-6 \times 10^{-6}$, but are extremely sensitive to the
normalization
of the power spectrum and cluster dynamics
\ref\CK  [\CK].
Therefore in flat $\Oo=1$ models, the quadratic Doppler effect
does not yield a measurable average (isotropic) \Cy\ parameter.
In \S IIIE, we calculate the fluctuations
in the \Cy\ parameter at arcminute
scales
and show that it too is negligible.  Note that the isotropic \Cy\ distortion
is an integral over optical depth $d\tau$, \ie it has a contribution from all
scatterings.  On the other hand,  correlations (\ie temperature fluctuations)
come from an integral over the visibility function $\exp(-\tau) d\tau$, \ie
they are
dominated by
the last scattering event.

The $v^2$ \Cy\ distortion
is non-negligible in the case of open universes.
In an open universe, velocities grow more slowly so that they are
still reasonably large at the scattering epoch.  Furthermore last scattering
happens at lower redshifts for baryon dominated open models.
In models with relatively low $\Omega_B$, predictions
for the quadratic Doppler effect are comparable to estimates of the
SZ effect, although still below the present constraints;
higher $\Omega_B$ models with an early photoionized IGM can already be
ruled out from
their larger thermal $y$-distortion
(see \eg \ref\GO [\GO]).

It is worthwhile
to note that such quadratic Doppler distortions
are the minimum required in the case of reionization
and unlike the SZ effect do not suffer from uncertainties in cluster dynamics
and ionization mechanisms.
\bigskip
\goodbreak
\centerline {\bf III. THE BRIGHTNESS EQUATION
AND SMALL SCALE ANISOTROPIES}
\medskip
\nobreak
\noindent A. GENERAL ARGUMENTS AND DEFINITIONS
\smallskip
In this
section, we will extend, refine and unify techniques developed in refs.
[\Vishniac, \Efstathiou, \KaiserB].  We employ the true angular
dependence of Compton scattering and treat the ${\cal O}(v^2)$
quadratic Doppler
effect as well as the ${\cal O}(v)$ and ${\cal O}(\vd)$ effects.
Furthermore, we find that
for $\Oo < 1$ models, there are significant corrections to the formalism of
Efstathiou [\Efstathiou].

Having shown that spectral distortions are
minimal and simply described by the \Cy\ parameter,
we will concentrate on temperature anisotropies, \ie
we shall be no longer be concerned with frequency dependence.
Therefore let us integrate
equation \eqnCIso\
over the spectrum to obtain
the evolution equation for the fractional energy
perturbation
to the spectrum $\D \equiv \delta \eden / \eden$.  This
yields the so called brightness equation
$$
\eqalign{
{\dot \D} + \G_i {\partial_i \D } &= \neav \sigma_T a
[1 + \delta(\X)] \Big\{ \D_0 - \D + {1 \over 2} P_2(\mu) \D_2 \cr
& \qquad + 4 \Bg \cdot \V -v^2 + 7 (\Bg \cdot \V)^2  \Big\},
\cr }
\eqn\eqnFull
$$
where the dot denotes differentiation with respect to conformal time,
and we have assumed that the cluster SZ effect is unimportant at the last
scattering epoch. The average SZ effect will produce an isotropic
\Cy\ distortion and have no effect on small scale anisotropies.

Here, we separate the
average electron density from the spatially dependent perturbation
as
$$
n_e = x_e(\E) (1-Y_p/2)n_B(\X,\E) = \neav(\E) [1 + \delta(\X,\E)],
\eqn\eqnNe
$$
where $Y_p$ is the primordial mass fraction in helium, $x_e$ is the
ionization fraction (assumed to be independent of position),
$n_B$ is the baryon number density and $\delta$
is the fractional overdensity in baryons.
The spatial dependence of the electron density distribution
couples with the ordinary source terms to provide
an additional effective source.  The coupling of density perturbations
to the velocity field
is in fact responsible for the
Vishniac effect.

Parenthetically, note that our definition of $\Delta$ does not
exactly conform with
the standard convention [\Efstathiou], where $\Delta =
\delta f (T/4)(\partial \bar f /\partial T) \approx
4(T_{\rm eff}(p)-T)/T$, with $\bar f$ a Planck
distribution at the average temperature $T$ and $T_{\rm eff}(p)$ the
effective temperature of the spectrum as a function of frequency.  Under
this definition, $\Delta$
retains spectral information.
We define $\Delta$ as the perturbation in the total energy density,  which
is explicitly independent of frequency.  The two definitions coincide
for a uniform shift in the temperature (\eg the linear Doppler effect),
which is the situation
 usually considered.   Care should therefore be taken in associating
our $\Delta$ with temperature distortions ({\it c.f.} equation \eqnTDy).

We can rewrite equation \eqnFull\ implicitly as
$$
{\dot \D} + \G_i {\partial_i \D } + \gnn
\D = \gnn S(\X,\Bg,\E),
\eqn\BSource
$$
where the conformal time derivative of the optical depth,
$\gnn = \neav \sigma_T a$, can be interpreted as the probability
of scattering in the interval $d\eta$.
Note that the source term $S(\X,\Bg,\E)$ includes the anisotropy suppression
$(\Delta_0)$ part.
The
other source terms, as we shall see, generate a small contribution to
the isotropic  fluctuations by scattering out of a given mode, thereby
partially
escaping the cancellation effect.  The isotropic part then contributes
by scattering back into the line of sight.  This of course does not
increase the anisotropy, only the inhomogeneity of the photon distribution.

To solve this equation, we transform to its Fourier
space analogue,
$$
\D(\K,\Bg,\E) = {1 \over V_x} \int d^3 x \D(\X,\Bg,\E) \exp(-i\K\cdot\X),
\eqn\eqnFConvention
$$
which has the well known solution [\Efstathiou],
$$
\D(\K,\Bg,\E) =
\int_0^\E S(\K,\Bg,\E')
 \gnnp
\exp [ik\mu(\E'-\E)] d\E',
\eqn\eqnBSoln
$$
where $\mu = \Bg \cdot \K / k$ and
$$
\gnnp = \gnn \exp \left[- \int_{\E'}^{\E} d\E'' {\dot \tau (\E'')}
\right] \equiv {d \over d\E} \exp[-\tau(\E,\E')]
\eqn\eqnVisibility
$$
is the so-called visibility function [note that $g(\E,\E)=\gnn$].
For the case of full ionization where $x_e=1$,
$$
\tau(z) = {\sigma_T H_0 \over 4\pi Gm_p} (1 - Y_p/2)
{\Omega_B\over \Oo^2}[2 - 3\Oo + (1 + \Oo z )^{1/2}(\Oo z + 3\Oo -2)],
\eqn\eqnOpticalz
$$
(where $m_p$ is the mass of the proton) or alternatively
$$
\tau(\E) = {\sigma_T H_0 \over 4\pi Gm_p} (1 - Y_p/2)
{\Omega_B \over \Oo^2} \left\{2 - 3\Oo  - (1-\Oo)^{3/2}
\coth \left[\E \over 2\Le \right]
\left( 3 - \coth^2 \left[ \E \over 2 \Le \right] \right)
\right\}, \qquad \Oo < 1
\eqn\eqnOpticale
$$
where $\Le = H_0^{-1} (1-\Oo)^{-1/2}$ is the curvature scale today
which merely translates our
convention for conformal time to the usual development angle
$\Psi = \E / \Le$.  For $\Oo=1$, the expression in braces should be
replaced by $\{(\E_0/\E)^3 -1\}$.

We are primarily interested in the solution at small wavelengths,
\ie $k\ \delta\E \gg 1$ where $\delta\E$ is the
conformal time ``thickness'' of
the last scattering surface, located at $\Ec$.  The precise
definition of $\Ec$ is somewhat arbitrary,
\eg $\Ec$ could
be defined as the epoch at which optical depth becomes unity
(we will employ this definition unless otherwise stated).  For late last
scattering $\delta\E \approx \Ec$; it is in this
sense that we mean the last scattering surface for reionization scenarios
is ``thick.''

Neglecting the evolution of the source term and the optical depth, the
integral over the plane wave for $\mu \ne 0$ in equation \eqnBSoln\
will in general cancel except for a small
portion of the order of a wavelength.
Therefore the resulting brightness fluctuations will be of order
$$
\D(\K,\Bg,\E) \approx S(\K,\Bg,\E) \tau_\lambda(\E), \qquad \mu \ne 0,
\eqn\eqnBScaling
$$
where $\tau_\lambda = (2\pi/k) \gnn$ is the optical depth across one
wavelength.
If the perturbation has many
wavelengths over the last scattering surface, $\tau_\lambda \ll 1$ since
the optical depth across the whole surface is of order unity.
Therefore, the brightness fluctuations are suppressed as compared
with the source terms.

The $\mu= 0$ mode escapes this cancellation.
Thus the brightness fluctuation
has a significant value only if the wavevector
$\K$ is perpendicular to the direction of the photon momentum, ${\bf \Bg}$.
The interpretation of this statement is simply
that if the wavevector were not strictly perpendicular to the photon
momentum vector, photons in a given direction would have scattered off
both crests and troughs of the perturbation, yielding net cancellation.

In the case of
the first order (linear Doppler)
 term, $S(\K, \Bg, \eta) = 4\Bg\cdot\V = 4 \mu v$ which vanishes
for the $\mu=0$ mode.
Therefore, as pointed out by Kaiser [\KaiserB], the first order
term is primarily cancelled up to a factor which involves the time evolution
of the perturbation.  It is important however to note that the general
cancellation argument holds for {\it any} source term,
while near perfect cancellation
will occur for $S(\K, \mu=0, \eta)=0$.

Having determined which perturbation modes give large contributions
to brightness fluctuations, we need to translate this information into
the angular correlation function of temperature fluctuations
on the sky today.  This is denoted by
$$
C(\theta,\sigma) = \left< {\D T \over T}(\Bg_1)
{\D T \over T}(\Bg_2) \right>_{sky},
\eqn\eqnCorr
$$
where the average is performed with fixed beam throw $\theta = \cos^{-1}
(\Bg_1 \cdot \Bg_2)$ and with Gaussian
beam-width $\sigma$ [$C(\theta,\sigma)$ is not
to be confused with
the Compton scattering collision terms $C(\X,\P)$].
First we must relate the
brightness distortion $\D$ to a fractional temperature perturbation.
For linear Doppler effects which change the temperature uniformly
across the spectrum and leave the spectrum a blackbody, $\D \approx 4 \D T/T$.
Quadratic and thermal Doppler effects create \Cy\ distortions such that
$\D \approx -2 (\D T/T)_{RJ}$ [see equation \eqnTDy].
To keep the expression applicable
to both sorts of distortions, we will write
$$
{\D T \over T} = {K \over 4} \D,
\eqn\eqnK
$$
where $K=1$ for the linear Doppler effect and $K= -2$ for quadratic
and thermal Doppler terms (\Cy\ distortions).

We will use the small scale approximation
given by Doroshkevich, Zel'dovich, \& Sunyaev
\ref\DZS [\DZS],
$$
C(\theta,\sigma) = {V_x \over 32\pi^2} \int_0^\infty k^2 dk
{K^2 \over 2} \int_{-1}^{1} d\mu
|\Delta(k,\mu,\eta)|^2
J_0 \left[kR_\E \theta (1 - \mu^2)^{1/2}\right]\exp
\left[-(kR_\E\sigma)^2(1-\mu^2)\right],
\eqn\eqnCorrGen
$$
where $R_\E$ translates comoving distance at epoch $\E$ to angle on the sky.
It is given by,
$$
R_\E(z) = {2 \over \Oo^2 H_0 (1+z)}\left\{\Oo z + (\Oo -2)[(1+ \Oo z)^{1/2}-1]
\right\},
\eqn\eqnRz
$$
or equivalently
$$
R_\E(\E) = {\cases {
\E_0 - \E, & $\Oo = 1$, \cr
\Le {\rm sinh}[(\E_0-\E)/\Le], & $\Oo < 1$. \cr
}}
\eqn\eqnReta
$$
Efstathiou [\Efstathiou] takes the
asymptotic form of equation \eqnReta\ for $\E_0 \gg \E$ [or
$(\Oo z)^{1/2} \gg 1$].  This is not
a good approximation for $\Oo < 1$ models in which the surface of
last scattering was relatively recent.  For example, if $\Oo = 0.2$
and $h=0.8$, then $z_* \approx 20$.

For  \Cy\ distortions, the correlation function [eqn. \eqnCorrGen]
will in principle measure the uniform
background $y$-distortion, estimated by equation \eqnYSecond,
 as well as fluctuations in the background.
If no spectral information is obtained, it is not possible to observe the
effects of the uniform distortion.  However, in practice, this is unlikely
to be
a concern; in small-scale experiments the uniform background is
always subtracted out.  For example, many experiments employ
beam switching to minimize noise, \eg a double beam experiment
measures $C(0,\sigma) - C(\theta,\sigma)$.
Whereas $C(\theta,\sigma)$ is sensitive to
fluctuations from all scales greater than $\sigma$, $C(0,\sigma)-C(\theta,
\sigma)$ is effectively only sensitive to angular scales between $\sigma$
and $\theta$.  Even maps generated by interferometry are only sensitive
to fluctuations below the size of the primary beam $\theta_{pb}$.

The cancellation arguments, which say that $|\D(k,\mu,\E)|$ is sharply peaked
at
$\mu=0$, allow us to set $\mu=0$ in the slowly varying terms, so that
$$
C(\theta,\sigma)
= \int_0^\infty Q(k)J_0 (kR_\E\theta)
\exp \left[-(kR_\E\sigma)^2 \right] dk/k ,
\eqn\eqnCorr
$$
where
$$
Q(k) = {K^2 V_x k^3 \over 64\pi^2} \int_{-1}^{1} |\D(k,\mu,\E)|^2 d\mu
\eqn\eqnPower
$$
expresses the power per logarithmic interval of the fluctuations.

Note that technically speaking since
the square of the brightness distortion enters into \eqnPower,
we need to
worry about cross terms between first and {\it third} order processes.
However, they are suppressed for two reasons.
Firstly, we are only interested in second order effects when the
first order term is suppressed.  The
cross terms are thereby also suppressed.
Secondly, as we shall see in part B,
for the linear Doppler effect,
the third order terms that couple with the
first order term are also themselves suppressed.
We will denote the order of the term by superscripts. For example,
$$
\eqalign {
\D &= \Da + \Db + \ldots, \cr
\delta &= \da + \db + \ldots, \cr
\V &= \V^{(1)} + \V^{(2)} + \ldots \cr
}
\eqn\eqnExpansion
$$
where superscript (1) refers to the values given by linear theory and
(2) refers to the mildly nonlinear next order correction, {\it etc.}

Finally, it should be mentioned that
equation \eqnCorrGen\
 is only strictly valid
for a thin last scattering surface.
It assumes that there is a one to one correspondence between angles and
length scales at last scattering given by equation \eqnReta.
It also assumes that
the phases of neighboring modes have been correlated by the scattering.
We expect errors from these approximation to be of order ${\delta \E
/ (\E_0-\Ec)}\approx \Ec/\E_0$ ($\simlt 0.2$ for realistic reionization
scenarios).
\goodbreak
\medskip
\noindent B. GEOMETRICAL CONSIDERATIONS
\smallskip
The physical picture becomes much clearer if we examine the geometrical
aspects of the linear Doppler effects.
Evolution under the linear Doppler and isotropic source terms
can be expressed as
$$
{\dot \D} + \G_i {\partial_i \D } + \gnn\D =
 \gnn [\D_0 + {1 \over 2} \D_2 P_2(\mu) + 4\Bg \cdot \Mc(\X)],
\eqn\eqnLDI
$$
where
$$
\Mc(\X) = [1 + \delta(\X)]\V(\X)
= {\bf v}^{(1)} + \delta^{(1)} {\bf v}^{(1)} +
{\bf v}^{(2)} + \ldots
\eqn\eqnMc
$$
is the so-called normalized matter current introduced by Vishniac [\Vishniac].
As discussed in part A  and shown explicitly below in part C,
the source terms involving $\Delta$ itself are small
compared to the Doppler terms, and thus higher order terms in $\D$ have
been dropped.  We can reinterpret
equation \eqnLDI\ as representing scattering off a uniformly
dense medium
of electrons with velocity $\Mc(\X)$.  The increase
in probability of scattering off overdense regions is then accounted for
by a rescaling of the velocity field.

For the linear Doppler terms, we can also think of the cancellation argument
[see discussion following equation \eqnBScaling]
as follows.  If the matter current has zero vorticity, the contribution
to the brightness fluctuation is approximately the line integral of the
gradient of a field.  The contribution is thus cancelled apart from
contributions from the end points.  This is equivalent to the cancellation
argument: if $\nabla \times \Mc(\X) = {\bf 0}$ then $\Mc(\K) \parallel \K$
and thus $S(k,\mu,\E) \propto \mu \approx 0$.

Now let us consider the matter current vorticity: $\nabla \times
\Mc(\X) = [1 + \delta(\X)] [\nabla \times \V(\X)]
- \V(\X) \times \nabla \delta(\X)$.
Since gravitational perturbations of a pressureless fluid are irrotational,
$\nabla \times \V(\X) = {\bf 0}$ to all orders.  However this is not true of
the
second term.
The vorticity of the $(v \delta)$ term
(and in second order theory this term alone)
does not vanish.
Although the velocity field is the gradient of a potential, it is
not the gradient of the local density field.
Small scale overdensities can move coherently with some
peculiar motion along the line of sight.  In these regions,
 there is
an increased probability of scattering and inducing a Doppler shift
in the background radiation temperature.  This will imprint an anisotropy
on the CMB at the scale of the scattering blobs.  Note that this is
an inherently non-linear process since we are coupling a large scale
(the scale of the velocity field) with a small scale (the scale of
the overdensity).  Recall that
the Fourier transform
of a product is a convolution, and
$$
\Mc_{\vd}(\K) = \Bv*\delta = \int_0^\infty d^3 k' \,\Bv(\Bk')\delta(\Bk-\Bk'),
$$
which shows the mode coupling explicitly.
Since $\V(\K') \parallel \K'$, this gives a net non-zero
 contribution perpendicular
to $\K$.  In other words, we can have
plane waves of the vector field $\Mc(\X)$, with
wave vector perpendicular to the line of sight, but with components
parallel to the line of sight.  This is exactly as required above:
there is no cancellation between crests and troughs for the
mode $\K \perp \Bg$,
but the Doppler effect does not vanish since $\Bg \cdot \Mc \ne 0$.

It is also useful to note that the perpendicular and parallel components
add in quadrature, \ie the cross terms between the Vishniac term and other
second order terms cancel
when we calculate $|\D(k,\mu,\E)|^2$, as we now show.
In all cases of interest, the source terms can be separated into
a term dependent on conformal time and terms dependent on the mode coupling,
$$
\Ec \gnnp S(\K,\Bg,\E') = A(k\Ec)G(r\equiv \E'/\Ec,\E)
\sum_{\K'} F(\Bg,\K,\K')\, \da(\K',\Ec) \,
\da(\K - \K',\Ec),
\eqn\eqnSep
$$
where we have written the conformal time ratio $\E'/\Ec$ as $r$.
Therefore employing equation  \eqnBSoln,
 we obtain the solution
$$
\Db(\K,\Bg,\E)
\approx A(k\Ec) G(k\mu\Ec) \exp(-ik\mu\E)
 \sum_{\K'} F(\Bg,\K,\K')\, \da(\K',\Ec)  \,
\da(\K - \K',\Ec) ,
\eqn\eqnSecGen
$$
where we
have approximated the integral over conformal time as a Fourier
transform by taking $\E \rightarrow \infty$ in both the limits of the
integral and
in $G$, \ie
$$
\eqalign{
\int_0^{\E} G(r,\E) \exp(ik\mu\Ec r)dr &\approx
\int_{0}^{\infty} G(r,\E=\infty) \exp(ik\mu\Ec r)dr \cr
&\approx G(k\mu\Ec). }
\eqn\eqnFourApprox
$$
Hence $G(k\mu\Ec)$ is the Fourier transform of $G(r,\E=\infty)$ with
$r\equiv \E'/\Ec$ as the Fourier
conjugate of $k\mu\Ec$.
Note that the quadratic
Doppler term also obeys equation \eqnSecGen\ as we shall see in part E.
On the other hand, the linear Doppler terms have the additional property that
$F(\Bg,\K,\K') = \Bg \cdot {\bf D}(\K,\K')$
with ${\bf D}(\K,\K')$ only a vector function of $\K$ and $\K'$.
Choosing our coordinate system so that $\K \parallel {\hat \X_3}$ and
$\Bg$ lies in the $x_1-x_3$ plane, we can rewrite
$$
\eqalign{
F(\Bg,\K,\K') &= a_\parallel^{(2)}(k,k',\cos\theta)\Bg \cdot \K +
a_\perp^{(2)}(k,k',\cos\theta) \Bg \cdot \K_\perp'  \cr
&= a_\parallel^{(2)}(k,k',\cos\theta)k \mu +
a_\perp^{(2)}(k,k',\cos\theta) (1 - \mu^2)^{1/2} k'\sin\theta
\cos\phi, \cr }
\eqn\eqnPP
$$
where $\K_\perp' = k'(\sin\theta \cos\phi, \sin\theta \sin\phi, 0)$ is the
component of $\K'$ perpendicular to $\K$, and $a_\parallel^{(2)}$,
$a_\perp^{(2)}$ are arbitrary
functions independent of the azimuthal angle $\phi$. Thus we can separate the
contributions into $\Db = \Db_{\parallel} + \Db_{\perp}$
which are due to $\Mc\parallel\K$ and $\Mc\perp\K$.

Cross terms such as
$$
\eqalign{
\D_\parallel^{*(2)} \Db_\perp & \propto
F_\parallel^*(\Bg,\K,\K') F_\perp(\Bg,\K,\K') \
|\da(\K',\Ec)|^2\  |\da(\K-\K',\Ec)|^2\ {k'}^2 dk' \sin\theta d\theta d\phi \cr
& \propto \int_0^{2\pi} \cos\phi d\phi = 0
\cr }
\eqn\eqnCross
$$
then vanish under integration over
the azimuthal angle $\phi$.  Here we have used the random phase hypothesis
to eliminate one of the sums over modes.  Of course, cross-terms between
first and second order vanish under this assumption as well.

Now consider the mixed first and third order term.  This can be
expressed as
$$
\D^{*(1)} \D^{(3)} \propto
\int_0^{2\pi} \left[a_\parallel^{(3)}(k,k',\cos\theta) k\mu
+ a_\perp^{(3)}(k,k',\cos\theta) (1 - \mu^2)^{1/2}k'
\sin\theta \cos\phi \right] d\phi.
\eqn\eqnThird
$$
The perpendicular part vanishes under the integration over azimuthal
angle.  However, the parallel part (for the linear Doppler effect)
is suppressed, as we can see, by the additional factor of $\mu$
in equation \eqnThird.
Hence mixed first and third order terms
are not only suppressed by the smallness of the first order term but also by a
corresponding suppression in the third order portion.
These
contributions are thus entirely negligible.

Since all relevant
contributions add in quadrature, we can calculate them independently.
In part C, we will calculate the first order Doppler and isotropic effects;
in part D, the Vishniac effect; and in part E the quadratic Doppler effect.
\goodbreak
\medskip
\noindent
C. FIRST ORDER DOPPLER AND ISOTROPIC EFFECTS
\smallskip
In Fourier transform space, the solution to the residual first order
effect is given by
$$
\Da(\K,\mu,\eta) = \int_0^\E \left[\Da_0 + {1 \over 2}P_2(\mu)\Da_2 +
4\mu \va \right]
 \gnnp
\exp [ik\mu(\E'-\E)] d\E'.
\eqn\eqnBFO
$$
Here we have  retained the $\D$ contributions to the source
in the first order integral despite
the suppression discussed in section part A, since there is a corresponding
suppression of the first order linear Doppler source term that makes
these two comparable.
Note furthermore that for the first order effect the source term is only a
function of $\Bg \cdot \K = k\mu$ and not of $\Bg$ in general.

Taking the zeroth moment with respect to $\mu$
of equation \eqnBFO, we obtain
$$
\Da_0(\K,\eta) = \int_0^\E d\E' \gnnp \left\{ \Da_0 j_0[k(\E-\E')] -
4i \va j_1[k(\E-\E')] + {1\over 2} \Da_2 j_2[k(\E-\E')] \right\},
\eqn\eqnZero
$$
where we have employed the identity
$$
j_n(z) = {1 \over 2} (-i)^n \int_{-1}^1 \exp(i\mu z ) P_n(\mu) d\mu.
\eqn\eqnSBessel
$$
Due to the oscillation of the spherical bessel functions,
in each case the integral only has strong contributions around $k(\E-\E')
\simlt 1$.  Since $k\Ec \gg 1$, this implies $\E \approx \E'$ (and of
course, this is just the cancellation argument in a different guise).
Therefore we can take the slowly varying quantities $g,
\Da_0,$ and $\va$ out of the integral.
As we have discussed in part A, the first term and third term are
suppressed by a factor
of order the optical depth across one wavelength of the perturbation.
We can perform the integral over the second term in equation~\eqnZero\
by noting
$$
\int_0^\infty j_n(z) dz = {\pi^{1/2} \over 2} {{\Gamma\left[ {1\over2}(n+1)
\right]} \over
{\Gamma\left[ {1\over2}(n+2)
\right]}},
\eqn\eqnSBesselI
$$
and therefore
$$
\Da_0 (\K,\E) \approx - {4 i\va(\K,\E)  \over k} \gnn.
\eqn\eqnDIso
$$

Taking the second moment of equation $\eqnBFO$, ignoring the $\Delta$
terms, and extracting the slowly varying quantities,
we obtain
$$
\Da_2(\K,\eta) \approx 4i\va(\K,\E) \gnn \int_0^\E d\E'
\left\{ {3 \over 5} j_3[k(\E-\E')] - {2 \over 5} j_1[k(\E-\E)] \right\}.
\eqn\eqnDQuad
$$
However, employing equation \eqnSBesselI,
we see that $\Da_2$ vanishes.  The
quadrupole term can thus be neglected as compared with the other
sources.

Now let us separate the time dependence of the perturbations.  If we
express the growth of density perturbations in linear theory by
$$
\da(\K,\E) = \drat (\E,\Ec) \da(\K,\Ec),
\eqn\eqnGrowth
$$
then the growth of velocity perturbations is
$$
\va(\K,\E) = i{\K \over k^2} \left[ {d \over d\E}\drat
(\E,\Ec) \right] \da(\K,\Ec),
\eqn\eqnGrowthv
$$
where $\drat(\E,\Ec) = D(\E)/D(\Ec)$ is the ratio of the growth factors
\ref\Weinberg [\Weinberg]:
$$
D(\E) \propto {\cases {
\E^2, & $\Oo = 1$, \cr
3 \sinh (\E/\Le) \left[\sinh (\E/\Le)
-  (\E/\Le) \right] / \left[ \cosh (\E /\Le) -1 \right]^2 - 2,
& $\Oo < 1$. \cr
}}
\eqn\eqnDgrowth
$$
Hence, for $\Oo=1$ or $\E / \Le
 \ll 1$ (alternatively $(\Oo z)^{1/2} \gg 1$),
$\drat(\E,\Ec) = (\E/\Ec)^2.$
Efstathiou [\Efstathiou] employs this approximation for $\Oo < 1$ models.
Again this is not a good approximation due to the fact that last scattering
is so recent in these models.  We use the correct growth
factors here.
Equation \eqnBFO\ can then be written as
$$
\Da(\K,\mu,\eta) =  {8\da(\K,\Ec) \over (k\Ec)^2} \e^{-ik\mu\E}
{\int_0^\eta
{\Ec^2 \over 2} \left\{ {d \over d\E'} \drat(\E',\Ec) \right\}
 [{\dot \tau (\E')} + ik\mu]
g(\eta,\eta') e^{ik\mu\eta'} d\eta'}.
\eqn\eqnDSolnFO
$$
Let us check that this solution is consistent with the assumption of
small $\Da_2$.
Taking the second moment of equation \eqnDSolnFO\ and employing \eqnSBessel,
we obtain
$$
\Da_2(\K,\E) =
{8\da(\K,\Ec) \over (k\Ec)^2} {\pi \over 4k}  {d \over d\E}
\left\{
{\Ec^2 \over 2} \left[ {d \over d\E} \drat (\E,\Ec) \right]
\gnn \right\}.
\eqn\eqnTwo
$$
We see that $\Da_2 / \Da_0 \approx {\cal O}(1 / k\Ec) \ll 1$ and
$\Da_2$ is indeed small
compared with the other source terms.

The integral over the conformal time in equation \eqnDSolnFO\
 is again approximately a
Fourier transform [\KaiserB]  of
$$
G_v(r,\E) = {1 \over 2}
\left[ {d \over dr} \drat(r\Ec,\Ec) \right]
 {\dot \tau(r\Ec)} g(\E,r\Ec) \Ec^2
 - {d \over dr} \left\{
{1 \over 2}
\left[ {d \over dr} \drat(r\Ec,\Ec) \right]
 g(\E,r\Ec) \Ec \right\}.
\eqn\eqnGv
$$
where the transform is defined in equation \eqnFourApprox.
Hence
$$
|\Da (k,\mu,\E)|^2 = {64 P(k,\Ec) \over (k\Ec)^4} |G_v(k\mu\Ec)|^2,
\eqn\eqnDsquared
$$
where we have implicitly used the ergodic and random phase hypotheses.
The power per logarithmic interval $Q(k)$ for the first order
Doppler effect is therefore
$$
Q_v(k)
= {2 \over \pi} {V_x \over \Ec^3} {I_{\E,v} \over (k\Ec)^2}
P(k,\Ec),
\eqn\eqnQv
$$
where
$$
\eqalign{
I_{\E, v} &\equiv  {1 \over 2\pi} \int_{-k\Ec}^{k\Ec} G_v(k\mu\Ec)d(k\mu\Ec)
\cr
&\approx \int_{0}^{\E_0/\Ec}  |G_v(r,\E_0)|^2 dr, \cr }
\eqn\eqnIeta
$$
and we have used Parseval's theorem to approximate
$I_{\E, v}$.   For full ionization with
$\Oo=1$ and $\Ec \ll \E_0$, $I_{\E,v} \approx 6.38$.
Efstathiou [\Efstathiou] employs this value in cases where the approximation
is not very good.  We shall see in part E that this can lead to a significant
correction.

One might worry about the appearance of $\Ec$ in the expression for $Q(k)$,
since
there is a certain arbitrariness in the definition of
the epoch of last scattering, \eg Efstathiou [\Efstathiou]  picks
$\tau(\Ec) = 1$,
whereas we might equally well have chosen the peak of the visibility function.
Nevertheless,
inspection of the growth factors in $P(k,\Ec)$ and $I_\E$ shows that
all dependence on $\Ec$ vanishes, as it must, since it has only been introduced
as a convenient normalization epoch.
\goodbreak
\medskip
\noindent
D. THE VISHNIAC EFFECT
\smallskip
Contributions to the brightness fluctuation due to the Vishniac term
$S_{\vd} (\X,\Bg,\E)=4\da(\X)[\Bg \cdot \V(\X)]$,
may be expressed as
$$
\Db_{\vd} (k,\mu,\E) = A_{\vd} (k\Ec) G_{\vd} (k\mu\Ec) \exp(-ik\mu\E)
 \sum_{\K'} F_{\vd} (\Bg,\K,\K') \da(\K',\Ec)
\da(\K - \K',\Ec) ,
\eqn\eqnVishniac
$$
where
$$
\eqalign{
A_{\vd}(k\Ec) &= {4i \over k\Ec}, \cr
G_{\vd}(r,\E)  &=
{1 \over 2} \left[ {d \over dr} \drat(r\Ec,\Ec) \right] \drat(r\Ec,\Ec)
g(\E,r\Ec)\Ec ,\cr
F_{\vd}(\Bg,\K,\K') &= \Bg \cdot \left( {\K' \over |\K'|^2} + {\K - \K'
\over {|\K - \K'|^2}} \right) k, \cr
}
\eqn\eqnVishFun
$$
with $G_{\vd}(k\mu\Ec)$ defined from $G_{\vd}(r,\E)$ as in equation
\eqnFourApprox.
The cancellation argument manifests itself here in that $G_{\vd}(k\mu\Ec)$
is
approximately the Fourier transform of a wide bell-shaped function
$G_{\vd}(r,\E_0)$.
If $k\Ec \gg 1$, this implies that only for $|\mu| \ll 1$ does $G_{\vd}
(k\mu\Ec)$
have strong contributions.  Therefore we can set $\Bg \perp \K$ for the
other slowly varying functions of $\mu$.  We are thus calculating
the perpendicular part $\Db_\perp$ alone as the Vishniac
term.

We find that
$$
Q_{\vd}(k) =
{1 \over 8\pi^3} {V_x^2 \over \Ec^6}
(k\Ec)^3
I_{\E, v\delta}
I_{k,v\delta}(k)
P^2(k,\Ec) ,
\eqn\eqnQvd
$$
where $I_{\E, \vd}$ is defined as in \eqnIeta\ in the obvious manner
and
$$
I_{k,\vd}(k) = \int_0^\infty dy \int_{-1}^1 d(\ct) {(1-\cst)(1-2y\,\ct )^2
\over (1+y^2-2y\,\ct )^2 } {P[k(1+y^2-2y\,\ct )^{1/2},\Ec] \over P(k,\Ec)}
{P[ky,\Ec] \over P[k,\Ec]}.
\eqn\eqnIkvd
$$
For full ionization
$I_{\E,v\delta} \approx 6.38$ when $\Oo=1$ and
$\Ec \ll \E_0$.  Again Efstathiou [\Efstathiou] uses this value for all cases.
In the next section, we shall see what corrections are necessary for the
general case.

Care must be taken when evaluating the mode coupling [equation \eqnIkvd].
The integrand has a divergence
on the boundary at $y=1, \cos\theta=1$.
As described in \S IVA, a resonance occurs
when
the density field $\da(\K')$
has a wavelength of $k \approx k'$ (which implies $y \approx 1$) and the
velocity field is coherent: $|\K-\K'| \ll k$ (\ie $\cos\theta
\approx 1$).
Furthermore, since $\K \perp \Bg$ to avoid cancellation,
 $(\K-\K') \parallel \Bg$
and yields a strong Doppler contribution.
The integral itself does not diverge however, since for reasonable
power spectra, the power on the largest scales tends toward zero.
Nonetheless,
it complicates the integral, since we can have significant coupling between
the largest and smallest scales.
 Thus even though for pure power laws
$I_{k, \vd}$ is independent of $k$, we expect that realistic power spectra
will deviate from this prediction.
This is especially true for
steeply red power spectra, where the large-scale velocity field is
determined far above
the scale of interest (\ie $k$) where the power spectrum is quite different.
Figure 1 displays the magnitude of this effect as a function of $k$ for
various choices of the small-scale power law $n$ and an arbitrary
large-scale cutoff in power at $k_{min}=0.001$ Mpc$^{-1}$.  For reference the
standard CDM model predicts $n=-3$, whereas BDM
models prefer $n= -1/2$ but may plausibly lie in the range $0 > n > -1$.
This figure should only be taken heuristically since we have not included
the proper large-scale power of a realistic power spectrum.
\medskip
\goodbreak
\noindent E. THE QUADRATIC DOPPLER EFFECT
\smallskip
The quadratic effect escapes cancellation since it is not directly
proportional to $\Bg \cdot \K = \mu$, but velocity perturbations are
smaller than density perturbations at small scales, since in linear theory
$v \propto \delta / k\E,$ if $\Oo=1$ or $(\Oo z)^{1/2} \gg 1$.
The effect is therefore
relatively
minimal at scales far below that of the horizon
as long as the mode coupling is not significantly
different from the Vishniac term.

The source term for the quadratic Doppler effect is
$$
S_{vv}(\X,\Bg,\E) =
  -\V^{(1)} \cdot \V^{(1)} + 7 [\Bg \cdot \V^{(1)}]^2 .
\eqn\eqnQDI
$$
In Fourier space, a product becomes a convolution so that
the solution can be written as
$$
\Db_{vv}(\K,\Bg,\E) = A_{vv} (k\Ec) G_{vv} (k\mu\Ec) \exp(-ik\mu\E)
 \sum_{\K'} F_{vv} (\Bg,\K,\K') \da(\K',\Ec)
\da(\K - \K',\Ec),
\eqn\eqnBvv
$$
where
$$
\eqalign{
A_{vv}(k\Ec) &= {4 \over (k\Ec)^2}, \cr
G_{vv}(r,\E) &=
\left\{ {1 \over 2} \left[
{d \over dr} \drat(r\Ec,\Ec) \right] \right\}^2
g(\E,r\Ec)\Ec, \cr
F_{vv}(\Bg,\K,\K') &=
 {{ - \K' \cdot (\K-\K') + 7 (\Bg \cdot \K') \Bg \cdot (\K-\K')}
\over |\K-\K'|^2 } {k^2 \over k'^2}. \cr
}
\eqn\eqnMode
$$
Again $G(k\mu\Ec)$ is approximately
the Fourier transform of a wide bell shaped function $G(r,\E_0)$,
and thus we can set $\Bg \perp \K$ in the other terms.
Notice that with this approximation
$$
F_{vv} = - {{ k k' \cos \theta + k'^2- 7 k'^2 \sin^2 \theta \cos^2 \phi}
\over {k^2 + k'^2 - 2kk'\cos\theta}} {k^2 \over k'^2},
$$
so the cross terms between this term and
the Vishniac term, $F_{\vd} \propto
\cos\phi$,  vanish under
integration over the azimuthal angle $\phi$.  Furthermore, the quadratic
Doppler effect
produces spectral distortions in addition to anisotropies
 so that in principle they
can be isolated from the linear Doppler effect.  In practice however,
we will find that the distortions from the quadratic Doppler effect
are too small to be presently observable.

The solution is
$$
Q_{vv}(k) = {1 \over \pi^3}
{V_x^2 \over \Ec^6} (k\Ec)
I_{\E, vv} I_{k, vv} (k)
P^2(k,\Ec),
\eqn\eqnPowervv
$$
where $I_{\E, vv}$ is defined as in \eqnIeta\ and
$$\eqalign{
I_{k, vv} &= \int_0^\infty dy \int_{-1}^{+1} d(\ct)
{(y-\ct)^2 - 7(1-\cst)(y-\ct)y + (147/8) (1-\cst)^2y^2 \over
(1+y^2-2y\,\ct )^2 } \cr
&\qquad \times {P[k(1+y^2-2y\ct)^{1/2},\Ec] \over P[k,\Ec]}
{P[ky,\Ec] \over P[k,\Ec]}.
\cr}
\eqn\eqnIb
$$
We have inserted the $K = -2$ factor from \eqnK\ under the assumption that
we are observing the Rayleigh-Jeans temperature distortions from this effect.
For full ionization,  $\Oo=1$, and $\Ec  \ll \E_0$, we find
 $I_{\E,vv} \approx 3\frac/2$.
In the general case, this value will change as does $I_{\E,v}$ and
$I_{\E,v\delta}$.  This is because $I_\E$ reflects the growth of perturbations
as compared with their value at $\Ec$, weighted by the probability of
scattering
at such an epoch.  Perturbations grow more slowly in an open
universe.  If
the integral is peaked above $\Ec$, as is $I_{\E,v\delta}$ and
$I_{\E,vv}$, the integral will decrease; if it is peaked below $\Ec$,
it will increase, as with $I_{\E,v}$.  There is also a small effect due
to cutting off the integral at the present time $\E_0$.  Figure 2 shows
the values of $I_\E$ as a function of $\Omega_B$ for $\Oo=\Omega_B$
and for $\Oo=1$.

Note that equation \eqnIb\
is very similar to the Vishniac $I_{k,\vd}$ integral.
Therefore the
$k$ dependence of $Q(k)$ will resemble the Vishniac power spectrum save
for the extra $1 / (k\Ec)^2$ term suppression at small scales.  Thus, unless
$$
{I_{k,vv} \over
I_{k,\vd}} \simgt {1 \over 8} (k\Ec)^2
{I_{\E,\vd} \over I_{\E,vv}},
\eqn\eqnRatio
$$
we expect the quadratic
Doppler effect to be smaller than the Vishniac effect at small angular
scales.
We calculate this ratio explicitly for a range of power laws, again
with an arbitrary large-scale cut off in power at $k_{min}=0.001$ Mpc$^{-1}$
(see Fig. 3).  In no case is the ratio sufficiently large to satisfy
equation \eqnRatio\ for $k\Ec \gg 1$.
At scales nearer the horizon at last scattering, the linear first
order Doppler effect is not cancelled, since there are only ${\cal O}(1)$
scattering blobs of this size across the last scattering surface.
Therefore, the quadratic Doppler effect plays no significant role
at these scales either.
\bigskip
\goodbreak
\centerline{\bf IV. CALCULATIONS FOR SPECIFIC COSMOLOGICAL MODELS}
\bigskip
\noindent
A. COLD DARK MATTER (CDM) SCENARIO
\smallskip
Recent comparisons of anisotropy limits from the South Pole '91 experiment
[\Gaier] with CDM normalized to COBE, have suggested that some CDM
models, especially those with high $\Omega_B$, may be ruled out,
\ref\Gorski \ref\Dodelson \eg [\Gorski, \Dodelson].
One possible resolution is to have early reionization to at least partially
erase the primordial anisotropies on degree scales.   If this were the
case, then secondary anisotropies would tend to be generated on smaller
scales, which we now consider in detail.

The CDM power spectrum is given by
\ref\EBW [\EBW]
$$
P(k,\E_0) =
{Ak \over \left\{ 1 + [ak + (bk)^{3/2} + (ck)^2 ]^\nu \right\}^{2/\nu}},
\eqn\eqnCDMpower
$$
where $a = 6.4 (\Gamma h)^{-1}$Mpc, $b = 3.0 (\Gamma h)^{-1}$Mpc,
$c = 1.7 (\Gamma h)^{-1}$Mpc and $\nu = 1.13$. For standard CDM, $\Gamma=0.5$.
The normalization
constant $A$ is given by COBE to be [\EBW]
$$
A = 5.2 \times 10^5 \left({ Q_{rms} \over 16\mu K} \right)^2 {(h^{-1}
{\rm Mpc})^4 \over V_x}.
\eqn\eqnCOBEA
$$
Figure~4 shows the radiation power spectrum $Q(k)$ for the various effects
in a universe that never recombines: $x_e(z)=1$.
For comparison,
we have plotted the primordial fluctuations from standard recombination
\ref\Holtzman [\Holtzman] in Fig.~4.
As expected, first order and quadratic Doppler terms are suppressed at
small scales.  Furthermore, the quadratic effect never dominates
on any scale.
The
result for the correlation function is plotted in Fig.~5.
Here the correlations are given for infinite beam resolution $\sigma=0$.
Note that the coherence scale is somewhat under an arcminute.

To obtain observable quantities, the formulae of Wilson \& Silk
\ref\WS [\WS]
can be employed, or we can note that for CDM on arcminute scales
and greater,
a double beam experiment essentially measures $C(0,\sigma)$, whereas
a triple beam experiment measures $3\frac/2 \ C(0,\sigma)$ for the Vishniac
effect, due to the small coherence scale.  In Fig.~6,
we show the predictions of $C(0,\sigma)$ for CDM with several values
of $\Omega_B$.

Recently, Tegmark \& Silk
\ref\TS [\TS]
have shown that reionization
may plausibly occur as early as $z_i \approx 50$.
Furthermore, they show that reionization
is sudden in realistic models, so that $x_e(z)$ is
essentially a step function at $z_i$.
For these models,
the optical depth will never reach unity
until the standard epoch of recombination,
and hence some primordial
anisotropies remain.  We will approximate this by employing the
cumulative visibility function,
$$
g_C(\eta_i) = \int_{\eta_i}^{\eta_0} g(\E_0,\E')d\E'
\eqn\eqnCumvis
$$
which tells us the fraction of the CMB photons which have been rescattered
since $z_i$ and
have thus suffered  suppression of primordial fluctuations.   In terms of
the correlation function, only a fraction
$f_{prim}(\E_i) = 1 - g_C^2(\E_i)$ of the original signal
remains after rescattering.

Of course,
the Vishniac effect decreases if the optical depth never reaches unity
on the new last scattering surface.
In this case, the epoch of new last scattering $\Ec$ is not the
conformal time when optical depth reaches unity.
We can however characterize this epoch by the maximum of the
visibility function.   Employing this definition, we can express the fraction
of the full effect for these models as $f_{v\delta}(\E_i)=Q_{v\delta}/
Q_{v\delta}^{\tau(\Ec)=1}.$
Figure~7 plots the weighting fraction $f(\E_i)$ for
the Vishniac effect and the primordial fluctuations.   This fraction should
be multiplied by the values in Fig.~4 to obtain the radiation power
spectrum in cases when optical depth never reaches unity on the new
surface of last scattering.
Notice that we still obtain a fairly large Vishniac effect even
when only a small fraction of photons have scattered and the primordial
fluctuations are nearly intact. For example,
if $\Omega_B=0.1$ and $z_i \approx 40$
then $f_{prim} \approx f_{\vd} \approx 0.9$, \ie
90\% of the primordial fluctuations remain {\it and} 90\% of the
maximum possible fluctuations from the Vishniac effect are generated.
 This is because the Vishniac effect
becomes stronger at later times due to the growth of the velocities.
Therefore even though only a small fraction of photons are rescattered
a relatively large anisotropy is imprinted on them.
So  late reionization
will {\it not} yield smaller CMB fluctuations on the arcminute
level.

We have also tested
a phenomenologically successful model inspired by CDM
that sets $\Gamma=0.2$.
This has the effect of suppressing the small-scale power with respect to
the large leading to an extremely small Vishniac effect (see Table 1).
\medskip
\goodbreak
\noindent
{ B. BARYONIC DARK MATTER (BDM) SCENARIO}
\smallskip
Baryonic dark matter models with isocurvature fluctuations contain
relatively large amounts of small-scale power
(for the power spectrum in these models, which are also referred to as
PIB or PBI models, see
\ref\Peebles  \ref\COP [\Peebles, \COP].
For this reason, primordial fluctuations from the
first order Doppler effect at standard recombination are overproduced
on the scale of the comoving particle horizon (\ie $\E$).
The angle subtended by the horizon at
redshift $z$ is
$$
2\, \sin (\theta_H/2) = {\Oo (\Oo z + 1)^{1/2} \over
\Oo z + (\Oo -2) [(\Oo z + 1)^{1/2} - 1]}.
\eqn\eqnHorizon
$$
At standard recombination $z \approx 1100$ and $\theta_H \approx 1^\circ$.
  Reionization is therefore necessary to
remove these degree scale fluctuations.
Generally, anisotropies will be erased up to the scale of the horizon
size at the epoch $z_*$ where optical depth reaches unity.
Secondary fluctuations
from the first order Doppler effect on the new last scattering surface
will of course be generated, and will peak near the
horizon size at $z_*$.
Since the
horizon scale is then larger than at standard recombination, the degree
scale constraints on CMB fluctuations can be avoided.
Primary fluctuations on degree scales will be erased and secondary
fluctuations will be smaller because they have lower amplitude {\it and}
because degree scales are well below their peak.

In Fig.~8, we
plot the angle subtended by the horizon at $\tau(\Ec)=1$.  Notice that
if $\Oo=0.1-0.2$ as suggested by some dynamical measurements, the horizon
at last scattering subtends $\sim 15^\circ$ today.  Therefore at degree
scales, there is significant cancellation as described above.  However,
for the case in which the total density includes matter that does not
contribute to the free electron density in the intergalactic medium (IGM),
\ie if baryons are hidden in compact objects or the ionization fraction
is less than unity, then the last scattering surface is further away.  In
low $\Oo$ universes, the angle subtended by the horizon at last
scattering is only marginally larger than that at standard recombination.
These models are therefore in danger of producing
secondary fluctuations on degree scales which are as large as
the primordial ones that
have been erased.

The relatively large amount of small-scale power in these models also leads
to large, and in fact observable, fluctuations at arcminute scales due
to the Vishniac effect.   Thus even models with a relatively late
last scattering epoch can be constrained.
We normalize the spectrum to the fractional mass fluctuation on
a scale of 8$h^{-1}$Mpc, \ie $\sigma_8 = 1$.
One complication arises however.  We have performed our calculations in
linear theory and therefore can only realistically
predict fluctuations on scales
larger than the non-linearity scale.   We define the non-linearity
scale as the value $k_{nl}$ for which
$$
{V_x \over {2\pi^2}} \int_0^{k_{nl}} P(k,\Ec) k^2 dk = 1,
\eqn\eqnNL
$$
corresponding to the scale on which $\delta \rho/\rho \simeq 1$.
The maximum anisotropy comes from taking
the linear power spectrum out to infinitely small scales.
The minimum anisotropy we expect comes from cutting off fluctuations
at scales smaller than the non-linearity scale, \ie setting  $Q(k>k_{nl})=0$.
Thus the true prediction lies somewhere between this minimum and maximum,
the uncertainty being caused
by our ignorance of nonlinear effects,
although we would expect the prediction using the cutoff at $k_{nl}$
to be more realistic.
Choosing $\Oo=0.2$ and $h=0.7$ as an example,
we plot the results of using this prescription
for various choices of the power law index $n$ in Fig.~9.  Note that for
$n > -1.5$ the fluctuations due to the Vishniac effect
 increase with decreasing angular size, and
therefore all experiments will effectively measure $C(0,\sigma)$.

Recent measurements from the VLA have yielded significant improvements
in the constraints on arcsecond fluctuations in the CMB.
\ref\Fomalont
Fomalont \etal\ [\Fomalont]
report $C(0,\sigma)^{1/2}\  \times 10^{5} <$
1.9, 2.1, 2.3, 4.0, 5.8, 7.2 for
$\sigma =$  $0.56',$\ $ 0.40',$\ $ 0.22',$\ $ 0.15',$\ $ 0.10'$ respectively.
Unfortunately, due to the uncertainty about effects
 below the non-linearity scale, the arcsecond
measurements cannot be used to constrain most of the BDM models,
and the larger angle measurements do not yet yield
an improvement over the Readhead \etal\
\ref\Readhead  [\Readhead]
 measurements at $\sigma = 0.78'$.
On the other hand,
a small improvement in the measurements at $\sigma > 0.2'$ would rule
out many
favored models, \eg $\Oo=0.2,\ h=0.8,\ n=0,\ x_e=1$ (see Table 1).

The most stringent constraints to date come from
the Australian Telescope Compact Array (ATCA)
which uses interferometry to make a sky map
and thus measure $C(0,\sigma)$.  Bounds are placed at
$\sigma = 0.87'$ on
$C(0,\sigma)^{1/2}$ amounting to $<0.9 \times 10^{5}$
[\Subrahmanyan].  This
correspondingly places limits on the BDM model parameters $\Oo$ and $n$
(see Fig.~10).  We have included a small contribution from the first
order term $Q_v$ under the prescription that as ATCA does not measure
structure above the primary beam size $\theta_{pb} = 2.5'$, there
should be a Gaussian cutoff in the power above these scales.
The tighter constraints on very low $\Oo$ models are actually from this
first order Doppler contribution.  For these models,
the last scattering surface is more distant, and the cancellation of the
first order Doppler effect becomes correspondingly less complete.
In general, whereas ${\cal O}(\vd)$ secondary anisotropies decrease with $z_*$,
${\cal O}(v)$ anisotropies actually {\it increase} with $z_*$.

Our constraints differ from Efstathiou [\Efstathiou], even accounting for
the new limits.   In particular, his scaling relations cannot be extended
to certain regions of parameter space.
The correction for the angle to distance relation $R_\E$ [equation~\eqnReta]
 tends to
increase the Vishniac effect for large $n$ and decrease the first order
term for small $n$.  The true conformal time integrals $I_\E$ (Fig.~2)
 tend to
decrease the Vishniac term and increase the first order
term for low $\Oo$.
The net effect is an increase in the predictions for low $\Oo$
and high $n$ and a decrease in the predictions for low $\Oo$ and
low $n$.   It is also interesting to note that the smaller scale experiments
do a better job of constraining high $n$ as opposed to low $n$.  This is
useful since the arcminute measurements tend to constrain low $n$
(see Table 1).

Notice that the $\Oo=0.2, h=0.8, n=-1/2, x_e=1$ model
often quoted in the
literature, \eg [\COP], is ruled out by
a small amount.  We can avoid such constraints by lowering the
density of free electrons in the intergalactic medium (IGM).
An {\it ad hoc} procedure would be to lower the ionization fraction
to a smaller constant value \eg $x_e=0.1$ [\COP].
But it is difficult to see
how the ionization fraction can be kept at a small but significant value
since the recombination time tends to be short.
In these models, the predictions of the Vishniac effect for
the ATCA experiment
are then below the observational limits (see Table 1).
On the other hand, we obtain an increase in the first order term.
Again, this is because the last scattering surface becomes distant and
cancellation is incomplete.  If we decrease the ionization fraction
substantially more than this, although primordial anisotropies will be erased,
new
ones of roughly the same size will be generated.
These models may overproduce fluctuations on degree scales (see Fig.~8).

A more physically motivated way of avoiding the constraint would be
to hide the baryons in black holes
[\GO].  Obviously,
this
has an effect quite similar to lowering the ionization fraction.   For
simplicity let us assume that the matter power spectrum is unchanged
from the $\Omega_B = \Oo$ case [recall that we have defined $\Omega_B$
as the fractional baryon density in the intergalactic medium (IGM)].
Fig.~11 shows how the combined
Vishniac and first order anisotropies are changed by hiding a fraction
of the baryons so that $\Omega_B<\Oo$.
The increase
in fluctuations for $\Omega_B/\Oo \ll 1$ is from the
first order term and, as described above, is quite significant.
For models in which the last scattering surface would already
have been distant (\eg $\Oo=0.1, h=0.5$) the increase in the first order term
is strong enough so that lowering the number of baryons in the IGM
actually increases the predicted fluctuations on arcminute scales.
Even in the favored
$\Oo=0.2, h=0.8$ model, hiding the baryons can only decrease the predictions
by a factor of $\sim 2$.  It seems that it may be possible to rule out
even these models given modest future improvements in observations.
\goodbreak
\bigskip
\centerline{\bf V. CONCLUSIONS}
\smallskip
We have shown that anisotropies on arcminute angular scales can provide
a powerful probe of the ionization history of the universe.  The
COBE DMR detection of large-scale angular anisotropies implies intermediate
angular scale anisotropies, (when combined with CDM or BDM power
spectra) that in the absence of reionization, exceed or are comparable to
observational limits.  Reionization is essential in BDM models, and
{\it may
be essential}  in CDM models, \eg [\Gorski, \Dodelson].
Moreover, efficient reionization {\it must} have occurred at $z>5$.
Reionization has sufficiently modest energetic requirements (at least
for photoionization) that it plausibly occurred when only a small fraction
of the universe was found in nonlinear structures [\TS].   This occurs
as early as $z \approx 1000$ in BDM and $z \approx 50$ in unbiased CDM
models.  While reionization helps suppress, or at least reduce,
primordial fluctuations
on degree scales due to the first order Doppler effect at standard
recombination, it inevitably regenerates secondary fluctuations on arcminute
scales.

We have considered all Compton effects to second order and have shown
explicitly that only
those of ${\cal O}(v)$ and ${\cal O}(\vd)$ play a significant role in
regeneration, \ie
no other higher order effects are important, and there are no hidden terms
to cancel the Vishniac term.
Furthermore, we have substantially improved the
approximations involved in calculating these effects in an open universe.
The ${\cal O}(v^2)$ effect however does set the minimal
spectral (\Cy) distortions in reionization scenarios.  For the CDM scenario,
all of these secondary effects predict fluctuations well below present
observations.  However, one should note that even in cases where the
primordial fluctuations are not efficiently erased due to relatively late
reionization, the Vishniac effect {\it still} generates fluctuations that are
of the same order as primordial fluctuations at arcminute scales.
For BDM models where early reionization is necessary, new measurements from
ATCA already exclude a significant region of
($\Oo,n$) parameter
space.  In fact, the standard model [\COP] of $\Oo=0.2,\, h=0.8$
with full ionization $x_e=1$ is
ruled out by a small amount.   Changing the ionization history can avoid
the present constraints, but it is difficult to decrease the secondary
fluctuations by a significant amount.  If one arranges the ionization
history to produce a small Vishniac effect, the first order Doppler effect
will be correspondingly increased.  We therefore find for both CDM and
BDM that changing the ionization history in what might be considered
physically well-motivated prescriptions
does not in fact greatly change the predictions for the secondary anisotropy.















\bigskip
\goodbreak
\bigskip
\centerline{\bf ACKNOWLEDGEMENTS}

We would like to express our debt to Scott Dodelson for
redirecting us toward properly evaluating the
collision integral.  We are grateful to Jim Peebles
and Renyue Cen for providing their BDM power spectra.
We acknowledge many useful discussions with
Roman Juszkiewicz, Naoshi Sugiyama, Max Tegmark, and Martin
White.  DS was supported by a
U.K.~SERC/NATO Postdoctoral Research Fellowship and by the CfPA.
This work has been supported in part by grants from NSF and DOE.
\eject
\centerline{\bf APPENDIX A: COLLISIONLESS BOLTZMANN EQUATION}
\smallskip
Second order gravitational effects
play an important
 role when the first order {\it gravitational}
 effect vanishes, \eg if there are
no metric fluctuations on the last scattering surface as in the case
of late phase transition scenarios.   These effects
will primarily manifest
themselves on large scales.   Martinez-Gonzalez,  \etal\
\ref\MGSS [\MGSS]
have performed
the calculation for weakly non-linear effects due to second order
density perturbations.  Here we complete the program by deriving all
possible effects to second order in the metric perturbations.
Note however that recently Jaffe \etal\
\ref\Jaffe [\Jaffe] have shown
that scenarios in which there are no fluctuations on the last scattering
surface must involve strongly non-linear effects to generate the observed
gravitational potentials today.  They find that such models must generate CMB
fluctuations on the order of, or greater than,
the standard primordial fluctuations.
Therefore, even for this class of models,  weakly
nonlinear effects such as those derived in here may only play the
role of a small correction to the full gravitational effect.

We choose to work in the synchronous gauge, following
closely the notation of Peebles and Yu
\ref\PY [\PY].  In this gauge, perturbations
to the metric are expressed as
$$
g_{00} = 1, \quad g_{0i} = 0,
\quad g_{ij} = -a(t)^2 [\delta_{ij} - h_{ij}(x,t)],
\eqnoh{(A--1)}
$$
where greek indices run from 0--3 and latin indices run from 1--3.  Summation
over repeated indices is assumed throughout even when all indices are lowered.
The momentum of the photon will be denoted by $p^\mu$ and for convenience
we express,
$$
p_i = -p_0 a(t) e \gamma_i,
\eqnoh{(A--2)}
$$
where $\gamma_i$ are the direction cosines.  In order to satisfy $p^\mu p_\mu
= 0$,
$$
e^2 = 1 - h_{ij}\gamma_i\gamma_j,
\eqnoh{(A--3)}
$$
to first order.

 The equations of motion for the photons
are [\PY]
$$
{\deriv p_\alpha;t } = {1 \over 2}  \partial_\alpha g_{\mu\nu} {p^\mu p^\nu
\over p_0 }.
\eqnoh{(A--4)}
$$
The general expression for the collisionless Boltzmann equation is
$$
{\part f;t } + {\part f;x^i }{\deriv x^i;t }
+ {\part f;\gamma_i } {\deriv \gamma_i;t }
+ {\part f;p_0 }{\deriv p_0;t } = 0.
\eqnoh{(A--5)}
$$
We integrate this equation over energy to obtain the equation for
$\D$, the brightness.  Each term in the integral of (A--5) is now
discussed in turn.
\medskip
\goodbreak
\noindent {\bf 1. Redshift Term}:
\smallskip
$$
\eqalign{
 {4\pi\over \eden} \int p_0^3dp_0 \,
{\part f;p_0 }{\deriv p_0;t } &=
4{1 \over a}{da \over dt}(1 + \Da + \Db)  - 2\G_i\G_j{\part h_{ij};t }
- 2\G_i\G_j {\part h_{ij}^{(2)};t }  \cr
& \qquad + \left\{- 2\G_i\G_j \Da - 4h_{kj}\G_k\G_i
+ 2h_{kl}\G_i\G_j\G_k\G_l \right\}
{\part h_{ij};t }, \cr }
\eqnoh{(A--6)}
$$
where $\eden$ is the average  energy density of the photons.
The first term represents the universal redshift due to the expansion.
Note that the brightness fluctuation is independent of the redshift factor
since all photons redshift in the same manner.  Thus, the term from the
explicit time dependence of the spectrum will cancel the first term
(see 4.).
The other terms represent gravitational redshifts due to perturbations
of the metric.
\medskip
\goodbreak
{\noindent \bf 2. Anisotropy Term}:
\smallskip
Anisotropy is generated as a first order gravitational effect due to
gravitational redshifts from overdense regions, \ie the Sachs-Wolfe effect
\ref\SW [\SW].
Therefore
$\partial f / \partial \gamma_i$ has a contribution to first order in
the metric perturbations.  Gravitational lensing, $d\gamma_i/dt$, is
also a first order effect.  Thus the whole term is second order:
$$
{4\pi\over \eden} \int p_0^3dp_0  \,
{\part f;\gamma_i } {\deriv \gamma_i;t } =
{\part \Da;\G_i } {1\over 2}\left\{ {\part h_{jk}; t } \G_i\G_j\G_k
 + {1\over a}\partial_i h_{jk} \G_j\G_k \right\}.
\eqnoh{(A--7)}
$$
\medskip
\goodbreak
{\noindent \bf 3. Inhomogeneity Term} :
\smallskip
$$
{4\pi\over \eden} \int p_0^3dp_0  \,
{\part f;x^i }{\deriv x^i;t } =
{\G_i\over a} \left\{ {\part \Da;x^i }
+ {\part \Db;x^i } \right\}
+ {1 \over a} {\part \Da;x^i }
\left[h_{ij} \G_j -  {1\over 2}h_{jk}\G_j\G_k\G_i
\right].
\eqnoh{(A--8)}
$$
This term represents the effects of a spatially dependent perturbation in
the photon energy density.  The first term has no dependence on the
metric fluctuations and is the flat space approximation used in
equation \eqnFull.  The second term represents corrections due to the
mass shell constraint, equations (A--2, A--3).
\medskip
\goodbreak
{\noindent \bf 4. Explicit Time Dependence Term} :
\smallskip
$$
{4\pi\over \eden} \int p_0^3dp_0 \,
{\part f;t }  =
-4{1\over a}{ da \over dt}\left( 1 + \Da + \Db \right)
+ {\part \Da;t }
+ {\part \Db;t }.
\eqnoh{(A--9)}
$$
The first term represents the uniform redshift of the spectrum.
It cancels with the energy redshift leaving
no effect on temperature perturbations.
\bigskip
\goodbreak
{\noindent \bf Complete First Order Expression:}
\smallskip
Summing the above, we arrive at
$$
{\part \Da;t } + {\G_i \over a} {\part \Da;x^i } =
  2\G_i\G_j {\part h_{ij};t },
\eqnoh{(A--10)}
$$
to first order.
This term represents the gravitational redshift of the photon due to
the metric fluctuations and gives the conventional Sachs-Wolfe effect
\ref\LSS [\LSS].
\bigskip
{\noindent \bf Complete Second Order Expression:}
\smallskip
In second order,
$$
\eqalign{
{\part \Db;t } + {\G_i \over a} {\part \Db;x^i }  &=
 2\G_i\G_j {\part h_{ij}^{(2)};t }
+  \left\{ 2\G_i\G_j \Da + 4h_{kj}\G_k\G_i
- 2h_{kl}\G_i\G_j\G_k\G_l \right\}
{\part h_{ij};t }. \cr
& \qquad
- {\part \Da;\G_i } {1\over 2}\left\{ {\part h_{jk}; t } \G_i\G_j\G_k
 + {1\over a}\partial_i h_{jk} \G_j\G_k \right\}
- {1\over a} \left\{{\part \Da;x^i }
\left[h_{ij} \G_j -  {1\over 2}h_{jk}\G_j\G_k\G_i\right] \right\} .
\cr
}
\eqnoh{(A--11)}
$$
This expression gives the second order Sachs-Wolfe effect and gravitational
lensing effects.
\eject
\centerline{\bf APPENDIX B: COMPTON COLLISION TERM}
\smallskip
Here we present the general expansion of equation~\eqnCGen\ to second order,
and give the explicit sources in the collision term of equation \eqnCImplicit.
First we must express the
matrix element in the frame of the radiation background:
$$
\eqalign{
|M|^2 &= (4\pi)^2 \alpha^2 2 \Bigg\{ \opc - 2\bomb \bigg[
{\pq \over mp} + {\ppq  \over mp'} \bigg] +
(1-\cosb)^2 {p^2 \over m^2} \cr
& \qquad +  {q^2 \over m^2} \bomb
+ (1-\cosb)(1-3\cosb)
\left[ {\pq \over mp} + {\ppq \over m p'} \right]^2 \cr
& \qquad
+ 2\bomb {(\pq)(\ppq) \over {m^2} {pp'}} \Bigg\}
+ h.o., \cr	}
\eqnoh{(B--1)}
$$
where higher order ({\it h.o.}) represents third
and higher order in $v$ and $p/m$.
The following identities are very useful for the calculation.
Expansion to second order in energy transfer can be handled in a quite
compact way by denoting it as an expansion
in $\delta p = q-q'$
of the delta function for
energy conservation:
$$
\eqalign{
\delta(p+q-p'-q') &= \del + {1 \over 2m} \left\{
{2 \pmp\cdot\Q }
+ {\pmp^2} \right\}
\ddel  \cr
&\qquad + {1 \over 8m^2}\left\{
{2 \pmp\cdot\Q }
+ {\pmp^2} \right\}^2
\dddel + h.o., }
\eqnoh{(B--2)}
$$
which is of course defined and justified by integration by parts.
Integrals over the electron distribution function are trivial,
$$
\eqalign{
\int {d^3 \Q \over (2\pi)^3} g(\Q) &= n_e, \cr
\int {d^3 \Q \over (2\pi)^3} q_i g(\Q) &= mv_i n_e, \cr
\int {d^3 \Q \over (2\pi)^3} q_i q_j g(\Q) &= m^2v_i v_j n_e +
mT_e \delta_{ij} n_e.\cr
}
\eqnoh{(B--3)}
$$
Now we
substitute all this into the general collision equation \eqnCGen\ and
obtain the explicit expressions for the terms of equation \eqnCImplicit,
$$
\eqalign{
C_0 =& \del \opc \fmf \cr
C_v =& \Bigg\{ \ddel \opc \V \cdot \pmp \cr
& \qquad - \del 2\bomb {\left[
{\pv \over p} + {\ppv \over p'} \right]} \Bigg\} \fmf \cr
C_{vv} =& {1 \over 2} \dddel \opc [\V \cdot \pmp]^2 \fmf \cr
& - \ddel 2\bomb \left[{\pv \over p}+ {\ppv \over p'}\right]
\V \cdot \pmp \fmf \cr
& + \del \Bigg\{ -(1-2\cosb+3\cos^2\beta)v^2
+ 2\bomb {(\pv)(\ppv) \over {p p'}}  \cr
& \qquad  +
(1-\cosb)(1-3\cosb)
\left[ {\pv \over p} + {\ppv \over p'} \right]^2
\Bigg\} \fmf\cr
C_{p/m} =& -\ddel \opc {{\pmp}^2 \over {2m}} \ffff \cr
C_{T_e/m} =& \Bigg\{ \dddel \opc {{\pmp}^2 \over 2}
- \ddel 2\cos\beta (1 - \cos^2\beta) (p-p') \cr
& + \del [4\cos^3 \beta - 9\cos^2\beta -1]
\Bigg\}
 {T_e \over m} \fmf, \cr }
\eqnoh{(B--4)}
$$
and the higher order terms,
$$
\eqalign{
C_{vp/m} =&
- \dddel \opc {\pmp}^2 {{\V \cdot \pmp} \over 2m} \ffff \cr
& + \ddel \Bigg\{ 2\cos\beta (1 - \cos^2\beta) (p-p')
{{\V \cdot \pmp} \over m}
\fof \cr
& \qquad + \bomb
{\pmp}^2 \left({\pv \over mp} + {\ppv \over mp'} \right) \ffff
\Bigg\} \cr
& + \del \Bigg\{ {{\V \cdot \pmp} \over m}
\left[
-\opc \fmf +
2 (1 - 2\cosb + 3\cos^2\beta)
\fof \right] \cr
& \qquad - 2\bomb \left[
\left( {p \over p'} -\cosb \right) {\ppv \over m}
- \left( {p' \over p} -\cosb \right) {\pv \over m}\right] \fof \Bigg\}\cr
C_{(p/m)^2} =&
\dddel \opc {{\pmp}^4 \over 2m^2}
\left[ {1 \over 4} \fmf + \fof \right] \cr
&  -\ddel 2\cosb (1-\cos^2\beta) (p-p') {{\pmp}^2 \over 2m^2} \fof \cr
& + \del
\Bigg\{ (1-\cosb)^2 \left({p \over m}\right)^2 \fmf
+ \opc {{\pmp}^2 \over 2m^2} \ffff \cr
& \qquad -  2\bomb{(p-p'\cosb)(p'-p\cosb) \over
m^2} \fof \cr
& \qquad  - [1-2\cos\beta+3\cos^2\beta] {{\pmp}^2 \over m^2} \fof
\Bigg\}
\cr
}
\eqnoh{(B--5)}
$$
where we have written
$$
\eqalign{
\fmf =& f(\X,\P')-f(\X,\P) \cr
\ffff =& f(\X,\P) + 2f(\X,\P)f(\X,\P') + f(\X,\P') \cr
\fof =& f(\X,\P')[1+f(\X,\P)] \cr
}
\eqnoh{(B--6)}
$$
for compactness.
\np
\noindent{\bf FIGURE CAPTIONS}
\smallskip
\noindent
FIG-1.  Mode coupling integral for the Vishniac effect ($v\delta$)
[see equation~\eqnIkvd]
 for various power laws [$P(k) \propto k^n$ for large $k$]
with a large-scale cutoff at $k_{min}=0.001$.
Note that for steep power spectra the integral never approaches a constant.
This represents a correction to the approximations in ref. [\Efstathiou].

\smallskip
\noindent
FIG-2.  The conformal time integrals $I_\E$ [see equation~\eqnIeta]
for the
${\cal O}(v)$,
${\cal O}(\vd)$, and
${\cal O}(vv)$ effects
for various models.  Note
that for $\Oo \ll 1$ the values depart significantly from the asymptotic
$\Oo=1,\ \Omega_B \ll 1$ result.  Again this significantly improves
the approximations of ref. [\Efstathiou].

\smallskip
\noindent
FIG-3.  Ratio of mode coupling integrals, \ie
$I_k$ of the quadratic Doppler effect ($vv$) to that of
the Vishniac effect ($v\delta$).   For no power law does the ratio become
sufficiently large to make the quadratic Doppler term significant
compared with the Vishniac term, at small scales.

\smallskip
\noindent
FIG-4.  CDM radiation fluctuation power spectrum for the first order Doppler
($v$), Vishniac ($v\delta$), and quadratic Doppler ($vv$) effects.
Here $Q(k)$ is the power per logarithmic interval, \ie $C(0) =
\int Q(k)\, dk/k$.  The shape parameter $\Gamma$ [see equation~\eqnCDMpower]
is $0.5$ for standard CDM.
Note
that the ($vv$) term is never significant: $(v)$ dominates at large
scales and ($\vd$) dominates at small scales.

\smallskip
\noindent
FIG-5.  CDM predictions for a double beam experiment with infinite resolution.
The primordial fluctuations for standard recombination are normalized to
COBE [\COBE].  Note that at sub-arcminute scales the Vishniac ($\vd$)
term dominates over the first order Doppler term ($v$) and can
be larger than the primordial fluctuations.  Also note that the
coherence scale of the Vishniac effect for CDM is somewhat under an
arcminute.

\smallskip
\noindent
FIG-6.	CDM predictions
of the Vishniac effect for $C(0,\sigma)$ for various choices of $\Omega_B$.
Because the coherence scale is under an arcminute, experiments at
arcminute scales and greater will essentially only measure $C(0,\sigma)$.

\smallskip
\noindent
FIG-7.  Weighting fraction for CDM models with sudden reionization at $z_i$.
The full values of Figure 4 should be weighted by $f$ to determine fluctuations
for these ionization histories.  Notice that we still obtain a strong
Vishniac effect even in cases where the primordial fluctuations are only
partially erased.  At $z_i \approx 40$, we have approximately 90\% of both
the Vishniac and primordial fluctuations. Late reionization therefore
does not necessarily yield smaller fluctuations.

\smallskip
\noindent
FIG-8.  Angle subtended by the horizon at last scattering
(optical depth unity), \ie $\Ec$
[see equation~\eqnHorizon].
In models where $\Omega_B=\Oo$, the horizon size is sufficiently
large to escape degree scale constraints.  However, models which hide matter
in compact objects or have low ionization fraction will have horizon sizes
close to the degree scale.  We have plotted for illustrative purposes
a model in which 90\% of the baryons are in compact objects ($\Omega_B
= 0.1 \Oo$) or
alternatively ionized fraction $x_e=0.1$.
\smallskip
\noindent
FIG-9.  BDM predictions for $C(0,\sigma)$ from the Vishniac effect
with $\Oo=0.7, h=0.7, x_e=1.0$.
This model barely escapes constraints by ATCA.  Notice that at arcseconds
we are in the non-linear regime; uncertainties here greatly inhibit our
ability to constrain models.  We have plotted predicted fluctuations with
(heavy) and without (light) the cutoff at $k_{nl}$.  These should be
taken as lower and upper bounds.

\smallskip
\noindent
FIG-10.  The limit from the ATCA measurement of
$C(0,0.87')< 0.9 \times 10^{-5}$ places
significant constraints on BDM models with full ionization $x_e=1$.
The constraints are strong
functions of the Hubble constant;
here we show three representative values of $h$, the lowest and highest values
generally considered and the specific value ($h=0.8$) chosen by ref. [\COP].
The boxed region indicates the preferred values
of ref. [\COP].  Low $n$ and high $\Oo$ are excluded.
Note that other assumptions about $x_e$ would alter these limits.

\smallskip
\noindent
FIG-11. Combined first order and Vishniac predictions for BDM models with
a significant fraction of the baryons in compact objects which
do not contribute to the baryon
density in the intergalactic medium, $\Omega_B$.   The first order effect
dominates for low $\Omega_B$, since the last scattering surface is then at
such high redshift.
\np
{\noindent \bf TABLE 1}
\bigskip
\noindent
\vbox{ \vskip 20pt \centerline{
\vbox{ \offinterlineskip
%
%
\halign {
\vrule#
& \hfil#\hfil& \vrule#
& \hfil#\hfil& \vrule#
& \hfil#\hfil& \vrule#
& \hfil#\hfil& \vrule#
& \hfil#\hfil& \vrule#
& \hfil#\hfil& \vrule#
& \hfil#\hfil& \vrule#
& \hfil#\hfil& \vrule#
\cr
\noalign{\hrule}
height2pt
&\omit&
&\omit&
&\omit&
&\omit&
&\omit&
&\omit&
&\omit&
&\omit&
\cr
%
%
&\quad Model\quad&
&\quad $\Omega_B$ \quad&
&\quad $h$ \quad&
&\quad spectrum \quad&
&\quad $x_e$ \quad&
&\quad $\sigma=0.1'$ \quad&
&\quad $=0.5'$ \quad&
&\quad $=1.0'$ \quad&
\cr
height2pt
&\omit&
&\omit&
&\omit&
&\omit&
&\omit&
&\omit&
&\omit&
&\omit&
\cr \noalign{\hrule}
height2pt
&\omit&
&\omit&
&\omit&
&\omit&
&\omit&
&\omit&
&\omit&
&\omit&
\cr
%
%
& BDM && $0.2$ && 0.8 && $n=0$  && 1.0 && 2.61-11.50 && 1.17-1.19 && 0.48 &
\cr
height2pt
&\omit&
&\omit&
&\omit&
&\omit&
&\omit&
&\omit&
&\omit&
&\omit&
\cr \noalign{\hrule}
height2pt
&\omit&
&\omit&
&\omit&
&\omit&
&\omit&
&\omit&
&\omit&
&\omit&
\cr
& BDM && $0.1$ && 0.8 && $n=-0.5$  && 1.0 && 2.17-3.26 && 0.80 && 0.48 &
  \cr
height2pt
&\omit&
&\omit&
&\omit&
&\omit&
&\omit&
&\omit&
&\omit&
&\omit&
\cr \noalign{\hrule}
height2pt
&\omit&
&\omit&
&\omit&
&\omit&
&\omit&
&\omit&
&\omit&
&\omit&
\cr
& BDM && $0.1$ && 0.5 && $n=-0.5$  && 1.0 && 1.77-2.36 && 0.49 && 0.25 &
  \cr
height2pt
&\omit&
&\omit&
&\omit&
&\omit&
&\omit&
&\omit&
&\omit&
&\omit&
\cr \noalign{\hrule}
height2pt
&\omit&
&\omit&
&\omit&
&\omit&
&\omit&
&\omit&
&\omit&
&\omit&
\cr
& BDM && $0.2$ && 0.8 && $n=0$  && 0.1 && 1.23-1.35 && 0.13 && 0.05 &
  \cr
height2pt
&\omit&
&\omit&
&\omit&
&\omit&
&\omit&
&\omit&
&\omit&
&\omit&
\cr \noalign{\hrule}
height2pt
&\omit&
&\omit&
&\omit&
&\omit&
&\omit&
&\omit&
&\omit&
&\omit&
\cr
& BDM && $0.2$ && 0.8 && $n=-0.5$  && 0.1 && 0.83 && 0.18 && 0.09 &
  \cr
height2pt
&\omit&
&\omit&
&\omit&
&\omit&
&\omit&
&\omit&
&\omit&
&\omit&
\cr \noalign{\hrule}
height2pt
&\omit&
&\omit&
&\omit&
&\omit&
&\omit&
&\omit&
&\omit&
&\omit&
\cr
& CDM && $0.1$ && 0.5 && $\Gamma=0.5$  && 1.0 && 0.38 && 0.25 && 0.18 &
  \cr
height2pt
&\omit&
&\omit&
&\omit&
&\omit&
&\omit&
&\omit&
&\omit&
&\omit&
\cr \noalign{\hrule}
height2pt
&\omit&
&\omit&
&\omit&
&\omit&
&\omit&
&\omit&
&\omit&
&\omit&
\cr
& CDM && $0.1$ && 0.5 && $\Gamma=0.2$  && 1.0 && 0.04 && 0.03 && 0.03 &
  \cr
height2pt
&\omit&
&\omit&
&\omit&
&\omit&
&\omit&
&\omit&
&\omit&
&\omit&
\cr \noalign{\hrule}
}}} \vskip 10pt}

\goodbreak
\bigskip
\noindent{\bf TABLE CAPTION}
\smallskip
\noindent TABLE-1.
Predictions of $C(0,\sigma)\ (\times 10^{-5})$
for the Vishniac effect for various models which can be tested
by future experiments at the VLA, ATCA, and OVRO.
We have chosen three representative values of the beam smoothing
$\sigma$.
Ranges given in the 6th and 7th column represent the uncertainty in
predictions due to nonlinearity.
The observed signal will also depend on the
first order effect which in turn depends sensitively
on the nature of the experiment
and not on $C(0,\sigma)$ alone.
\np
{\bf \centerline{ REFERENCES}}
\smallskip
\refs[\HKR] C.J. Hogan, N. Kaiser, and M.J. Rees, Phil. Trans. R. Soc. Lond.
{\bf 307}, 97 (1982).

\refs[\COBE] G.F. Smoot \etal, Astrophys. J. Lett., {\bf 396}, L1 (1992).

\refs[\Gaier] T. Gaier \etal, Astrophys. J. Lett., {\bf 398}, L1 (1992).

\refs[\OV] J.P. Ostriker, and E.T. Vishniac, Astrophys. J., {\bf 353} ,372
(1987).

\refs[\Vishniac] E.T. Vishniac, Astrophys. J., {\bf 322}, 597 (1987).

\refs[\Efstathiou] G. Efstathiou,  Large Scale Motions in the Universe: A
Vatican Study Week, eds. Rubin, V.C. and Coyne, G.V., (Princeton University,
Princeton, 1988) pg. 299.

\refs[\Subrahmanyan] R. Subrahmanyan \etal, \MNRAS, (in press, 1993).

\refs[\BE] J.R. Bond and G. Efstathiou,  \MNRAS, {\bf 226},
665 (1987).

\refs[\KaiserA] N. Kaiser, \MNRAS, {\bf 202}, 1169 (1983).

\refs[\Bernstein] J. Bernstein, Relativistic Kinetic Theory,
(Cambridge University, 1988).  This approach was suggested to us by
S. Dodelson.

\refs[\MS] F. Mandl and G. Shaw, Quantum Field Theory (Wiley, New York, 1984)
pg. 157.

\refs[\ZIS] Ya. B. Zel'dovich, A.F. Illarionov, R.A. Sunyaev,  Sov. Phys.--
JETP, {\bf 33}, 643 (1972).

\refs[\KaiserB] N. Kaiser, Astrophys. J., {\bf 282}, 374  (1984).

\refs[\SZ] R.A. Sunyaev and Ya. B. Zel'dovich, Comments Astrophys.
Space Phys., {\bf 4}, 79 (1972).

\refs[\Kompaneets] A.S. Kompaneets, Sov. Phys.--JETP, {\bf 4}, 730 (1957).

\refs[\Bond] J.R. Bond, The Early Universe, eds. Unruh, W.G. and
Semenoff, G.W., D. Reidel, (Dordrecht, Boston, 1988) pg. 283.

\refs[\ZS]  Ya. B. Zel'dovich and R.A. Sunyaev, Ap\&SS, {\bf 4}, 301 (1969).

\refs[\BS] J. Bartlett and A. Stebbins, Astrophys. J. {\bf 371}, 8 (1991).

\refs[\Smoot] G.F. Smoot \etal, Astrophys. J. Lett., {\bf 371}, L1 (1991).

\refs[\Mather] J.C. Mather, \etal, Astrophys. J. Lett. (submitted, 1993).

\refs[\CK] S. Cole and N. Kaiser, M.N.R.A.S., {\bf 233}, 637 (1988);
J. Bartlett and J. Silk, Ap. J., (submitted 1993).

\refs[\GO] N.Y. Gnedin, and J.P. Ostriker, Astrophys. J.,
{\bf 400},1
(1992).

\refs[\DZS] A.G. Doroshkevich, Ya. B.  Zel'dovich, and R.A. Sunyaev,
 Sov. Astron., {\bf 22}, 523 (1978).

\refs[\Weinberg] S. Weinberg, Gravitation and Cosmology
(Wiley, New York, 1972).

\refs[\Gorski] K.M. Gorski, R. Stompor, and R. Juszkiewicz, Astrophys. J.,
(submitted, 1993).

\refs[\Dodelson] S. Dodelson and J.M. Jubas, Phys. Rev. Lett., {\bf 70}, 2224
(1993).

\refs[\EBW] G. Efstathiou, J.R. Bond, and S.D.M. White, \MNRAS,
{\bf 258}, 1p (1992).

\refs[\Holtzman] J. Holtzman, Astrophys. J. Supp., {\bf 71}, 1 (1989).
We have normalized to COBE [\COBE].

\refs[\WS] M.L. Wilson and J. Silk, Astrophys. J., {\bf 243}, 14 (1981).

\refs[\TS] M. Tegmark, and J. Silk,  Astrophys. J.,  (in press, 1993).

\refs[\Peebles] P.J.E. Peebles, Nature, {\bf 327}, 210 (1987).

\refs[\COP] R. Cen, J. Ostriker, P.J.E. Peebles, Astrophys. J.,
(submitted, 1993).

\refs[\Fomalont] E. Fomalont, \etal, Astrophys. J. {\bf 404}, 8 (1993).

\refs[\Readhead] A.C.S. Readhead \etal, Astrophys. J.,{\bf 384},396 (1989).

\refs[\MGSS] E. Martinez-Gonzalez, J.L. Sanz,
and J. Silk, Phys. Rev. D, {\bf 46}, 4193
(1992).

\refs[\Jaffe] A. Jaffe, A. Stebbins, and J.A. Frieman,
Astrophys. J. (submitted, 1993).

\refs[\PY] P.J.E. Peebles, and J.T. Yu, Astrophys. J., {\bf 162}, 815 (1970).

\refs[\SW] R.K. Sachs and A.M. Wolfe, Astrophys. J., {\bf 147}, 73 (1967).

\refs[\LSS] P.J.E. Peebles, Large Scale Structure of the Universe,
(Princeton University, Princeton 1980).

\bye